\documentclass[aps,prx,reprint,superscriptaddress,amsmath]{revtex4-1}
\usepackage{lipsum}
\usepackage{graphicx}
\usepackage{subfigure}
\usepackage{textcomp}
\usepackage{setspace}

\begin{document}
\bibliographystyle{apsrev4-1} 

\title{Solution to the key problem of statistical physics\\ -- calculations of partition function of many-body systems}
\author{Bo-Yuan Ning}
\affiliation{Center for High Pressure Science $\&$ Technology Advanced Research, Shanghai, 202103, China}
\author{Le-Cheng Gong}
\affiliation{Institute of Modern Physics, Fudan University, Shanghai, 200433, China}
\affiliation{Applied Ion Beam Physics Laboratory, Fudan University, Shanghai, 200433, China}
\author{Tsu-Chien Weng}
\affiliation{Center for High Pressure Science $\&$ Technology Advanced Research, Shanghai, 202103, China}
\author{Xi-Jing Ning}
\email{xjning@fudan.edu.cn}
\affiliation{Institute of Modern Physics, Fudan University, Shanghai, 200433, China}
\affiliation{Applied Ion Beam Physics Laboratory, Fudan University, Shanghai, 200433, China}

\date{\today}

\begin{abstract}
The key problem of statistical physics standing over one hundred years
is how to exactly calculate the partition function (or free energy) of many-body interaction systems,
which severely hinders application of the theory for realistic systems.
Here we present a novel approach that works at least four orders faster than
state-of-the-art algorithms to the problem and can be applied to predict thermal properties of 
large molecules or macroscopic condensed matters via \emph{ab initio} calculations.
The method was demonstrated by C$_{60}$ molecules, 
solid and liquid copper (up to $\sim 600$GPa),
solid argon, graphene and silicene on substrate,
and the derived internal energy or pressure is in a good agreement with the results of 
vast molecular dynamics simulations in a temperature range up to $2500$K, 
achieving a precision at least one order higher than previous methods.
And, for the first time,
the realistic isochoric equation of state for solid argon was reproduced directly from the partition function.
\end{abstract}


\maketitle
\section{Introduction}
By the end of $19$th century,
the born of statistical physics brought a promising prospect that
products of complex chemical reactions\cite{fe1} and all the thermodynamic properties of macroscopic systems,
such as equation of states (EOSs)\cite{eos1,eos2} and phase transitions\cite{kofke},
can be thoroughly predicted without empirical data by calculating the free energy (FE) or partition function (PF).
Nevertheless,
it was soon realized that such an implementation is impossible for large molecules and condensed matters\cite{pfconden}
because PF is in principle determined by all the possible microstates over entire phase space and
a $3N$-fold integral has to be solved to obtain the PF for a system consisting of $N$ particles,
which goes far beyond the capability of modern supercomputers if the standard algorithm of numerical integral is applied\cite{liquids}.
For instance,
a rough calculation of PF for a C$_{60}$ molecule, involving a $180$-fold integral,
would cost at least $10^{100}$ years
by using even the fastest high-performance computing facility with $\sim10^{16}$ FP64 operations per second.

Alternative routes are resorted to sampling approaches, that is,
either following the formalism of molecular dynamics (MD) simulation to trace the trajectories by integrating Newton's equation of motion\cite{MD},
or,
in a manner of Monte Carlo (MC) way to explore the microstates with substantial contributions to PF by stochastic walk in phase space\cite{MC}.
Although time-average-based MD and ensemble-average-based MC algorithms have been developed over half a century
and proved to be impressive to gain mechanical properties\cite{mcmd},
when it comes to thermodynamic properties,
the long-standing problem in face of both schemes remains that a balance has to be reluctantly sought for between limited-length sampling and
the demanding requirement of ergodicity as the size of a system increases\cite{pes1}.
Progress has been made in evaluations of the relative difference of PF (or free energy)\cite{jarz1997,kofke15},
and more attentions are being paid to the density of state to calculate absolute PF\cite{wl1,*DOS,*nxj,*multicanonical,*transitionmatrix}.
Nested sampling may be state-of-the-art technique\cite{nestsample1},
which aims at uniformly sampling a series of fixed fractions partitioned by potential energies (PEs) in configurational space\cite{nestsample3,*nestsample7}
and has been applied in several systems described by empirical potentials\cite{nestsample5,nestsample10,do1,nestsample9,nestsample4,do3,do4,do2,nestsample11,nestsample12,nestsample13}.
Despite of its improved $2$-$3$ orders of computational efficiency,
NS can hardly work with \emph{ab initio} calculations
because of too much computational cost,
and even if pairwise interaction potentials are employed,
the affordable systems are limited to a scale of hundreds particles.

In this work, instead of tackling PF in the fashion of sampling,
we established a direct integral approach (DIA) to
solve the $3N$-fold integral on the basis of \emph{reinterpretation} of original sense of integral, 
which works at least four orders faster than NS algorithm. 
Validations of the method were made in the systems including
C$_{60}$ clusters, condensed copper, solid argon, graphene and silicene on substrate,
where the internal energies or EOSs obtained by DIA were compared to MD simulations,
and for solid argon,
the computed isochoric EOSs under high pressure zone were directly compared to experiments.
Excellent agreements confirmed the accuracy of DIA.
Ultrahigh efficiency of DIA paves a way to calculate the FE of large molecules and macroscopic condensed matters with \emph{ab initio} computations.
\section{Direct Integral Approach to Partition Function}
\label{sec.method}
\begin{figure}
\centering
\includegraphics[width=3.5 in,height=2in]{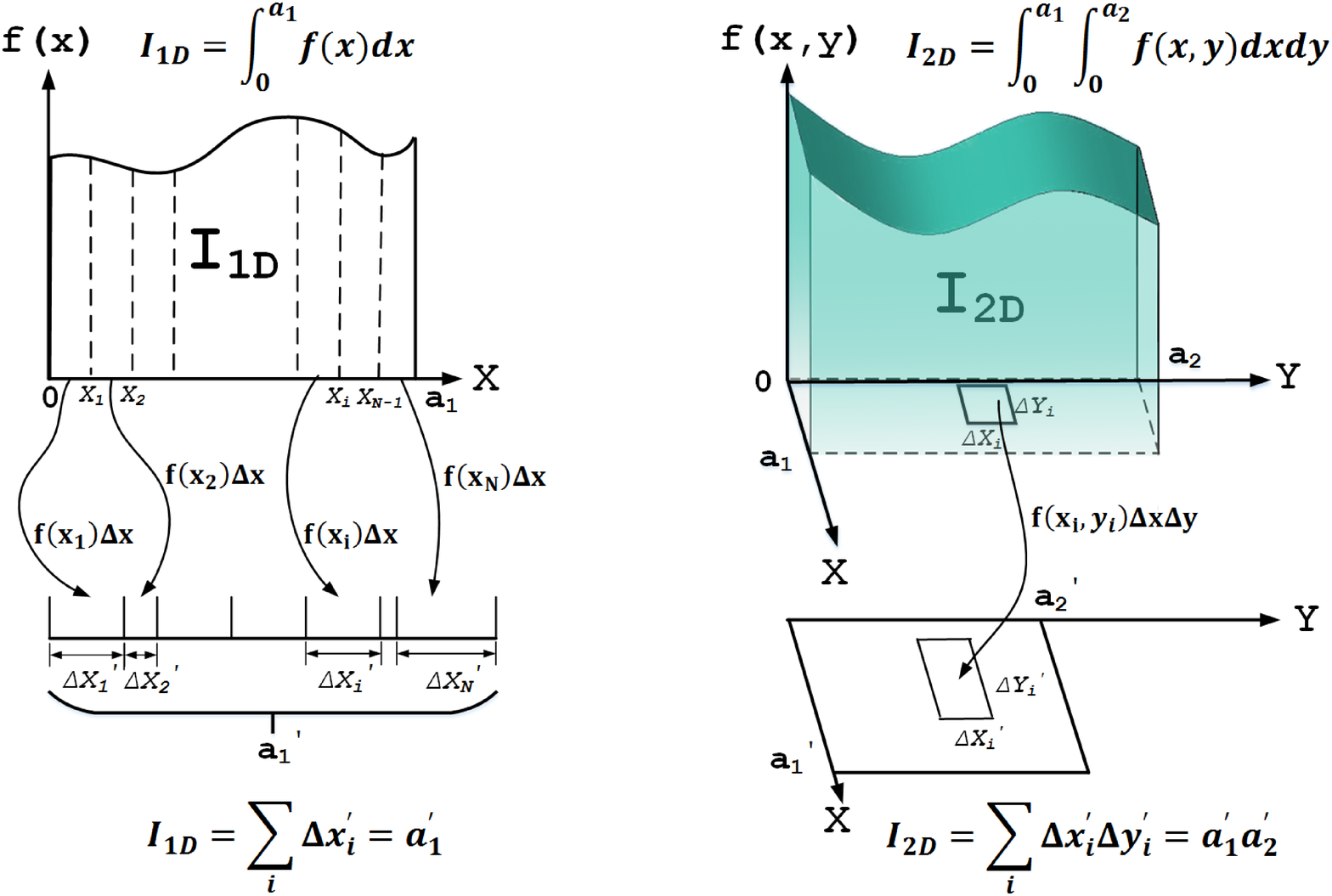}
\caption
{A one-fold integral can be interpreted as an effective length (left) and a two-fold integral equals to an effective area (right).
\label{fig1}
}
\end{figure}

The original sense of one-fold ($1D$) integral $I_{1D}=\int_0^{a_1} f(x)dx$ is interpreted as
the sum of infinite number of rectangles with area $A_i=f(x_i)\Delta x$, and
$\displaystyle I_{1D}=\lim_{\Delta x\to 0}\sum_iA_i$.
Here, we interpret the integral from a different angle:
The length of the $1D$ element $\Delta x$ at $x_i$ is modulated by $f(x_i)$ to be a new length element $\Delta x'_i=f(x_i)\Delta x$
and $\displaystyle I_{1D}=\sum_i\Delta x'_i$.
In other words,
the $1$D integral is mapped to a summation of length elements instead of area elements
and equals to an effective length of $a_1$ (see left in Fig.\ref{fig1}).
Similarly, a two-fold integral $ I_{2D}=\int_0^{a_1}\int_0^{a_2}dxdyf(x,y)$ equals to an effective area of $a_1\cdot a_2$
because the area element $ds=dxdy$ is enlarged (or shrunk) by $f(x,y)$
giving rise to an effective area element $ds'=f(x,y)dxdy$ (see right in Fig.\ref{fig1}).
Followed by this notion,
an $N$-fold integral $I_{ND}=\int_0^{a_1}\int_0^{a_2}\ldots\int_0^{a_N}dq_1dq_2\ldots dq_Nf(q_1,q_2\ldots q_N)$
equals to an effective volume of $a_1\cdot a_2\ldots a_N$.

When the integrand $f(q_1,q_2\ldots q_N)$ is in a form of $\exp[-U(q_1,q_2\ldots q_N)]$ with
$U(q_1,q_2\ldots q_N)$ being positive definite within the integral domain and
having minimum at the origin ($U(0)=0$),
the effective length of $a_i$ is defined as
\begin{equation}
\label{eq1}
a'_i=\int_0^{a_i}\exp[-U(0\ldots q_i\ldots 0)]dq_i,\quad (i=1,2\ldots N)
\end{equation}
and the effective volume approximates to a product $\displaystyle \prod_{i=1}^Na'_i$
(see proof in Supplementary Information), i.e.,
\begin{equation}
\label{eq2}
\displaystyle I_{ND}\simeq\prod_{i=1}^Na'_i.
\end{equation}

Now consider the configurational integral~(CI) in PF ($\mathcal{Z}$) for a continuum system consisting of $N$ particles at a given temperature $T$ (see details in Supplementary Information),
\begin{equation}
\mathcal Q=\int dq^{3N}\exp[-\beta U(q^{3N})],
\label{eq3}
\end{equation}
where $\beta=1/k_BT$ with $k_B$ the Boltzmann constant,
$q^{3N}=\{q_1,q_2 \ldots q_{3N}\}$ the Cartesian coordinates of particles
and $U(q^{3N})$ the potential function.
Although the integrand is of the same form as required by Eq.(\ref{eq2}),
it may not be positive definite or have no minimum at the origin ($q^{3N}=0$).
Letting the set $Q^{3N}=\{Q_{1},Q_{2} \ldots Q_{3N}\}$ be the coordinates of particles in state of the lowest potential energy $U_0$,
we may introduce a function
\begin{eqnarray}
\label{eq4}
U'(q'^{3N})=U(q^{3N})-U_0,
\end{eqnarray}
where $q'_i=q_i-Q_i$.
By inserting Eq.(\ref{eq4}) into Eq.(\ref{eq3}),
we obtain
\begin{equation}
\mathcal{Q}=e^{-\beta U_0}\int dq'^{3N}\exp[-\beta U'(q'^{3N})].
\label{eq5}
\end{equation}
Clearly, $U'(q'^{3N})$ is positive definite within all the integral domain and
has minimum at the origin ($U'(0)=0$).
According to Eq.(\ref{eq2}),
the integral in Eq.(\ref{eq5}) equals to an effective $3N$-fold volume,
\begin{equation}
\label{eq6}
\mathcal Q=e^{-\beta U_0}\prod_{i=1}^{3N}\mathcal{L}_i,
\end{equation}
where the effective length $\mathcal{L}_i$ on the $i$th degree of freedom is defined as
\begin{equation}
\label{eq7}
\mathcal{L}_i=\int e^{-\beta U'(0\ldots q'_i\ldots 0)}dq'_i.
\end{equation}
In this way, the $3N$-fold integral is turned into one-fold integrals.

For some homogeneous systems with certain geometric symmetry,
such as perfect one-component crystals,
all the particles are equivalent and
$U'$ felt by one particle moving along $q'_x$ may be the same as the one along $q'_y$ (or $q'_z$).
In such a case,
Eq.(\ref{eq6}) turns into
\begin{equation}
\mathcal Q=e^{-\beta U_0}{\mathcal{L}}^{3N},
\label{eq8}
\end{equation}
where $\mathcal{L}$ is the effective length determined by Eq.(\ref{eq7}).
Otherwise, it is needed to calculate the effective length,
$\mathcal{L}_x$, $\mathcal{L}_y$, $\mathcal{L}_z$ by Eq.(\ref{eq7}) of an arbitrary particle,
and Eq.(\ref{eq6}) turns into
\begin{equation}
\label{eq9}
\mathcal Q=e^{-\beta U_0}(\mathcal{L}_x\mathcal{L}_y\mathcal{L}_z)^N.
\end{equation}
The procedure can be extended to systems composed of different particle species by
calculating the effective length of each species respectively.

For inhomogeneous systems,
such as defects or interfaces existed,
particles may be grouped into $M$ sets numbered by $I$ with each containing $N_I$ equivalent particles,
and CI becomes
\begin{equation}
\label{eq10}
\mathcal Q=e^{-\beta U_0}\prod_{I=1}^{M}\left[\mathcal V_I\right]^{N_I},
\end{equation}
where $\mathcal V_I=\int e^{-\beta U'(x)}dx\int e^{-\beta U'(y)}dy\int e^{-\beta U'(z)}dz$
denotes the effective volume of an arbitrary particle in the $I$th set with Cartesian coordinates $x$, $y$, $z$.

To implement DIA,
the first step is to find the most stable structure (MSS) of the system for determining $U_0$,
which can be accomplished in principle by several well-developed methods,
such as global optimizations\cite{glopt1,*glopt2,*glopt3,*glopt4} or dynamic damping\cite{lowestu2,lowestu1}.
Actually, the MSS for crystals can be immediately obtained by placing the particles right at the lattice sites.
Then, we move a particle along its one degree of freedom $q'_i$ to obtain $U'(0\ldots q'_i\ldots0)$
while its other degrees of freedom $q'_j$ and all the other particles are kept fixed.
Clearly, this is an easy task for \emph{ab initio} calculations and therefore,
the PF of a $N$-particle system can be obtained
even if we have no knowledge about the analytical expression of potential function $U(q^{3N})$,
which is usually hard to be constructed precisely for realistic systems composed of more than one kind of particles.

For testing the DIA,
we may perform \emph{ab initio} calculations of $U'(0\ldots q'_i\ldots 0)$ on some realistic systems
and compare the derived results with related experiments.
However,
results from first principle calculations may be strongly dependent on the specific algorithms,
such as different types of exchange-correlation functions in density functional theory,
and the experimental data are usually insufficient for extensive comparisons.
In such cases,
even if the results derived from the PF are in good agreement with the experimental data,
it would be yet doubted of the accuracy of DIA.
In order to have a stringent test,
empirical potentials were used in the computations of PF,
and the derived results were compared to the MD simulations using the same potentials to
see if there exist some deficiency in DIA.

\section{Results and discussions}
\subsection{For isomers of C$_{60}$ molecule}
\label{sec.res.c60}
\begin{figure}
\begin{minipage}[t]{0.3\linewidth}
\centering
\includegraphics[width=1in,height=1.9in]{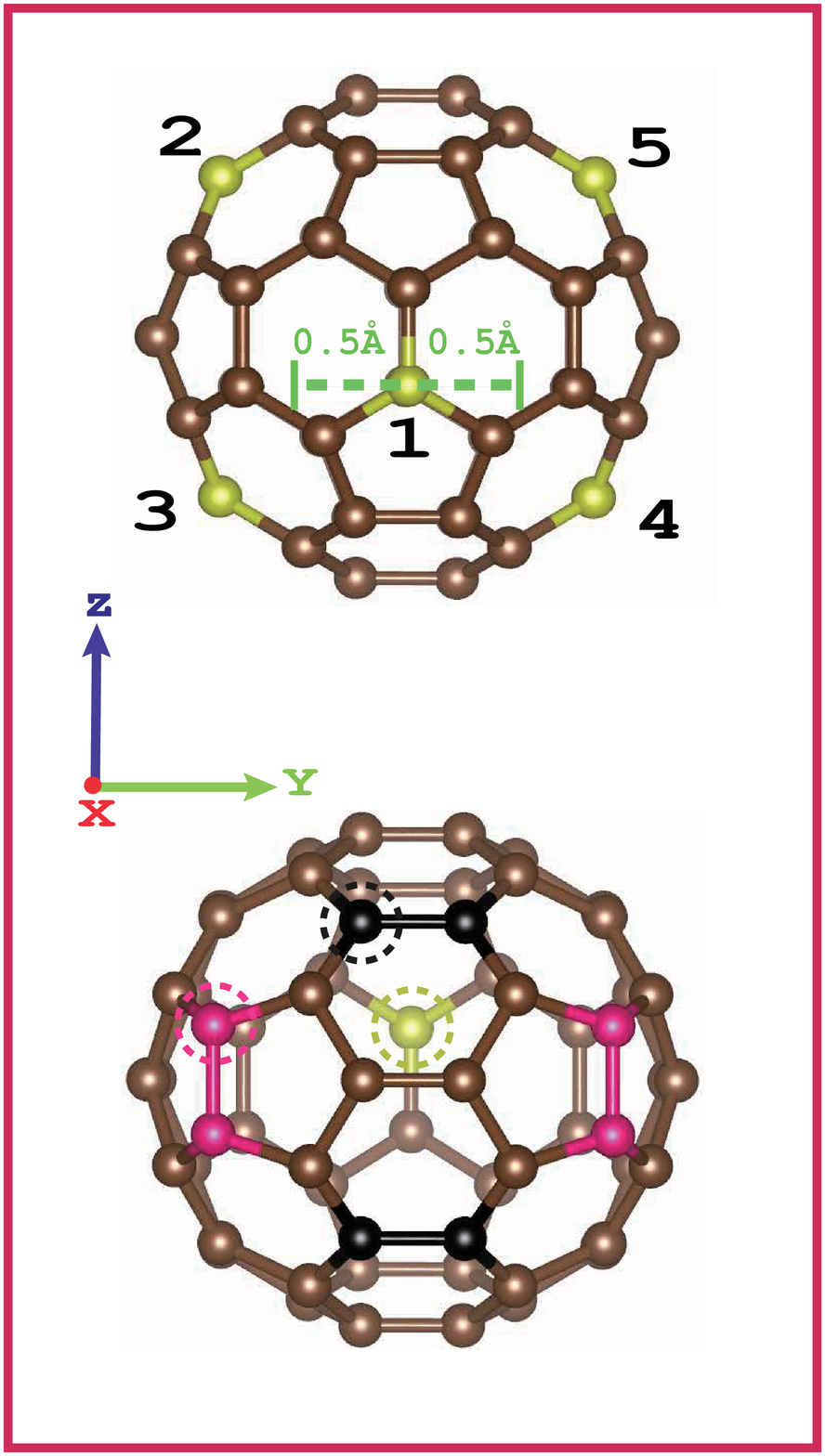}
\label{c60a}
\centerline{(a)}
\end{minipage}
\begin{minipage}[t]{0.62\linewidth}
\includegraphics[width=2.4in,height=1.9in]{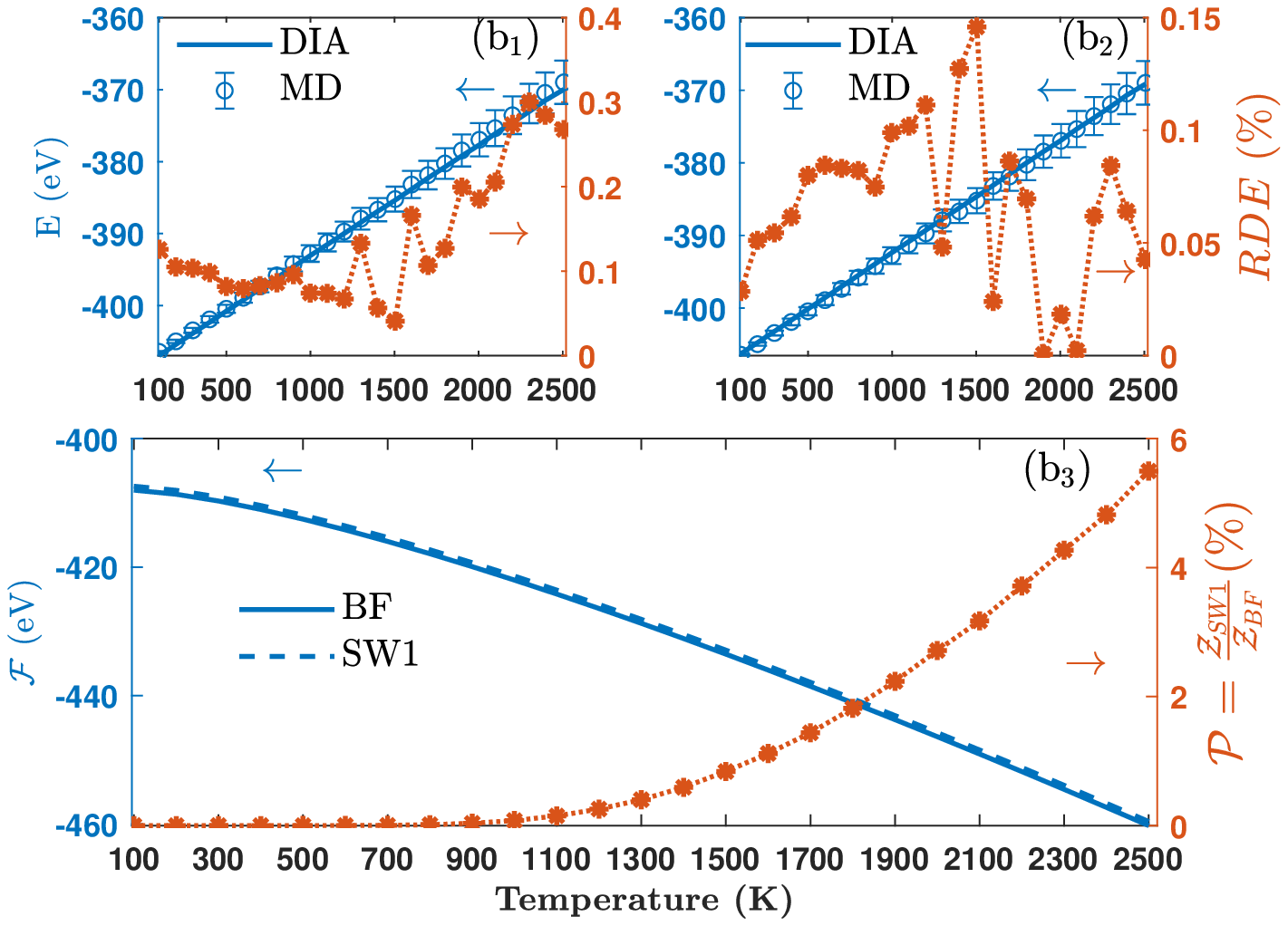}
\label{c60b}
\centerline{(b)}
\end{minipage}
\caption
{
DIA for C$_{60}$ clusters. 
(a) MSSs of BF molecule (top) and SW$1$ isomer (bottom). 
(b) Internal energies derived from $\mathcal{Z}_{BF}$ (b$_1$) and $\mathcal{Z}_{SW1}$ (b$_2$) were compared to those obtained by MD simulations. 
Error bars represent the standard deviations of the MD simulations. 
The FEs of the two molecules and the surviving probability of SW$1$ isomer are shown in (b$_3$).
\label{figc60}
}
\end{figure}

For a cluster consisting of $60$ carbon atoms, 
the most appealing impression may be the discovery of buckminsterfullerene (BF) with a football-cage structure\cite{c60}.
Following investigations revealed that, besides the lowest potential-energy BF, 
there exists a bunch of C$_{60}$ isomers, 
such as those introduced by the Stone-Wales (SW) rearrangement\cite{c60sw1}. 
For a long time, 
researches\cite{c60sw2,c60sw3,c60swnxj} have been attempting to answer the very questions 
whether the isomers are able to survive in realistic systems at finite temperatures 
and what the probability relative to BF is. 
The most reasonable argument should be the ratio 
of the PF of isomers ($\mathcal{Z}_{ISM}$) to that of BF ($\mathcal{Z}_{BF}$), 
but to our best knowledge, 
no such theoretical works have been tried out. 
Among all the isomers, 
the stack-$1$ SW (SW$1$) isomer (see bottom in Fig.\ref{figc60}(a)) has the lowest PE, 
which is $\sim0.6$eV higher than that of BF when using many-body Brenner potential\cite{c60swnxj}. 
Here DIA was applied to calculate $\mathcal{P}=\mathcal{Z}_{SW1}/\mathcal{Z}_{BF}$ to 
determine the relative surviving probability of SW$1$ isomer in a temperature range from $100$K to $2500$K.

To implement DIA, 
we firstly determined the MSS of the BF and the SW$1$ isomer by the Polak-Ribiere conjugate gradient algorithm. 
For the MSS of BF as shown in the top panel of Fig.\ref{figc60}(a), 
each atom is shared by two hexagons and one pentagon, 
indicating that all the atoms are geometrically equivalent and 
thus only one atom is needed for the calculation of $\mathcal{Z}_{BF}$. 
For an arbitrary atom, 
the felt PEs along its three Cartesian coordinates are obviously not the same, 
so $\mathcal{L}_x$, $\mathcal{L}_y$ and $\mathcal{L}_z$ has to be calculated respectively 
and Eq.(\ref{eq9}) was applied. 
The selected atom was moved step by step with an interval of $0.0001$ {\AA} along its Cartesian $X$-axis 
(or $Y$, $Z$-axis, 0.5 {\AA} in positive and negative direction) to 
obtain the corresponding potential-energy curves, $U'(X)$ (or $U'(Y)$, $U'(Z)$), 
during which, 
the other two coordinates and all the other atoms were kept fixed (see details in Supplementary Information). 
Fig.\ref{figc60}(b$_1$) shows the internal energy $E$ of BF derived from the PF, $E=-\partial\ln\mathcal{Z}/\partial\beta$, 
where the atom labeled No.1 in the top panel of Fig.\ref{figc60}(b$_1$) was selected to calculate $\mathcal{Z}_{BF}$. 
As a comparison, 
MD simulations were performed to calculate the internal energy
and the Nose-Hoover algorithm~\cite{nosehoover} for canonical ensemble was employed with a time step $0.1$ fs. 
The systems were allowed to relax 20 ps initially and continued to run for another 50 ps,
during which averages of $E$ were recorded in every 10 fs.
All the computations concerning the C$_{60}$ molecules were carried out in 
the Large-scale Atomic/Molecular Massively Parallel Simulator software package~\cite{lammps}, 
and the many-body Brenner potential\cite{brenner} was selected to characterize the interactions between carbon atoms. 
As we can see, 
$E$ obtained from $\mathcal{Z_{BF}}$ are in an excellent agreement with those from MD simulations and 
the relative difference of $E$, RDE($=|(E_{PF}-E_{MD})/E_{MD}|$), is $<0.4\%$ up to $2500$K, 
which verifies the accuracy of the PF obtained by DIA. 
It should be noted that, 
based on our derivations, 
DIA is \emph{independent} of how the directions of the Cartesian coordinates are chosen. 
To test this, 
we arbitrarily selected four more atoms, labeled No.$2$-$4$ in the top panel of Fig.\ref{figc60}(a), 
to calculate the $\mathcal{Z}_{BF}$ respectively. 
Although the potential-energy curves for different atoms are not the same, i.e., $U'_1(X)$ differs a lot from $U'_4(X)$, 
the calculated internal energies based on the five atoms are the same (see detailed data listed in Supplementary Information).  

For the MSS of SW$1$ isomer shown in the bottom panel of Fig.\ref{figc60}(a), 
atoms were divided into three groups, 
where the first group is the atoms shared by two hexagons and one pentagon (colored in brown), 
the second one is the atoms shared by one hexagon and two pentagons (colored in magenta), 
and the last one is the atoms shared by three hexagons (colored in black).
Accordingly, 
Eq.(\ref{eq10}) was used to calculate the $\mathcal{Z}_{SW1}$, and, 
an arbitrary atom in each group, surrounded by dashed lines in bottom panel of Fig.\ref{figc60}(a), 
was selected to obtain the corresponding potential-energy curves respectively. 
The selected atoms were moved $0.5$ {\AA} in positive and negative direction, 
and $10^4$ PEs were recorded to obtain the corresponding potential-energy curves. 
Fig.\ref{figc60}(b$_2$) shows the comparison of the internal energy derived from $\mathcal{Z}_{SW1}$ and MD simulations, 
where the RDE is $<0.15\%$ within the whole temperature range. 
To test the universality of DIA, 
we additionally chose other different atoms in the three groups to calculate the $\mathcal{Z}_{SW1}$ respectively 
and the obtained results are the same as those shown in Fig.\ref{figc60}(b$_2$) (see detailed data listed in Supplementary Information). 

Fig.\ref{figc60}(b$_3$) shows the obtained FEs of BF and SW$1$ isomer. 
Although the FE of BF remains smaller than that of SW$1$ isomer within the whole temperature range,  
the difference is quite small, 
which makes it difficult to conclude whether BF is the more stable one than the isomer or not. 
On the other hand, 
an unambiguous picture was presented by calculating the ratio of PF between these two clusters.  
Apparently, SW$1$ isomer has negligible probability to survive at low temperature zone compared to BF, 
while, for $T>1300$K, 
the surviving probability of SW$1$ prominently increases and ends up to $\sim 6\%$ at $2500$K, 
exhibiting that there is a considerable chance of the isomer to form at higher temperature zone. 
Such a finding qualitatively agrees with previous results by $40$ns-long MD simulations\cite{c60swnxj} 
but the exact value of $\mathcal{P}$ differs from each other, 
which calls for further clarifications by future experiments.

\subsection{For solid and liquid Copper}
\label{sec.res.cu}
Considering the huge computational cost of the MD simulations rather than DIA,
the number of copper (Cu) atoms in our model is limited to $4000$,
which were confined in a cubic box with periodic boundary condition (PBC) applied.
The tight-binding (TB) potential\cite{tb} was employed to describe the interatomic interactions.
For solid Cu,
the MSS was found by arranging the atoms at the FCC lattice sites.
In consideration of the Fm-3m symmetry of FCC lattice,
the Cartesian $Z$-axis of atoms is set to [$001$] direction
so that the potential $U'$ felt by an atom moving along the $X$-axis (or $Y$-axis) is the same as the one along the $Z$-axis.
To obtain $U'(Z)$,
the $Z$ coordinate of the geometry-center atom was changed step by step with its $X$ and $Y$ coordinates fixed to record the potential energies,
during which all the other atoms stay fixed as well.
For the liquid system,
the atoms were heated up to $2.5\times10^4$K in MD simulation to generate a uniform distribution and
a damped trajectory method\cite{lowestu2} was used to determine the MSS and potential energy $U_0$ (see details in Supplementary Information).
Different from the cases in crystal Cu,
the potential $U'$ felt by a liquid atom moving along the $X$-, $Y$- or $Z$-axis may not be the same,
so $U'(x)$, $U'(y)$ and $U'(z)$ were calculated respectively and Eq.(\ref{eq9}) was applied to obtain the CI.

Common procedures for MD simulations of a canonical ensemble\cite{lowestu1} was employed to produce the internal energy ($E$) and pressure ($P$)
of the systems contacted with a thermal bath at given temperatures,
and the Verlet algorithm\cite{verlet} was employed for integrating the equations of motion
with time step $0.1$ fs and $0.01$ fs for solid and liquid respectively.
We first fully relaxed the systems for $10^5$ steps, and
ran another $10^5$ (or $10^6$) steps to record the values of $E$ and $P$ for solid (or liquid).
Attentions have to be paid that
the commonly used Virial equations\cite{liquids} are inaccurate to compute pressure in the case of many-body potential with PBC applied\cite{vt1,vt2}, and we employed the method proposed by Tsai\cite{vt3}
that considers stress and momentum flux across an area for conducting the statistic of $P$.
\begin{figure}
\begin{minipage}[t]{0.481\linewidth}
\includegraphics[width=1.68in,height=1.7in]{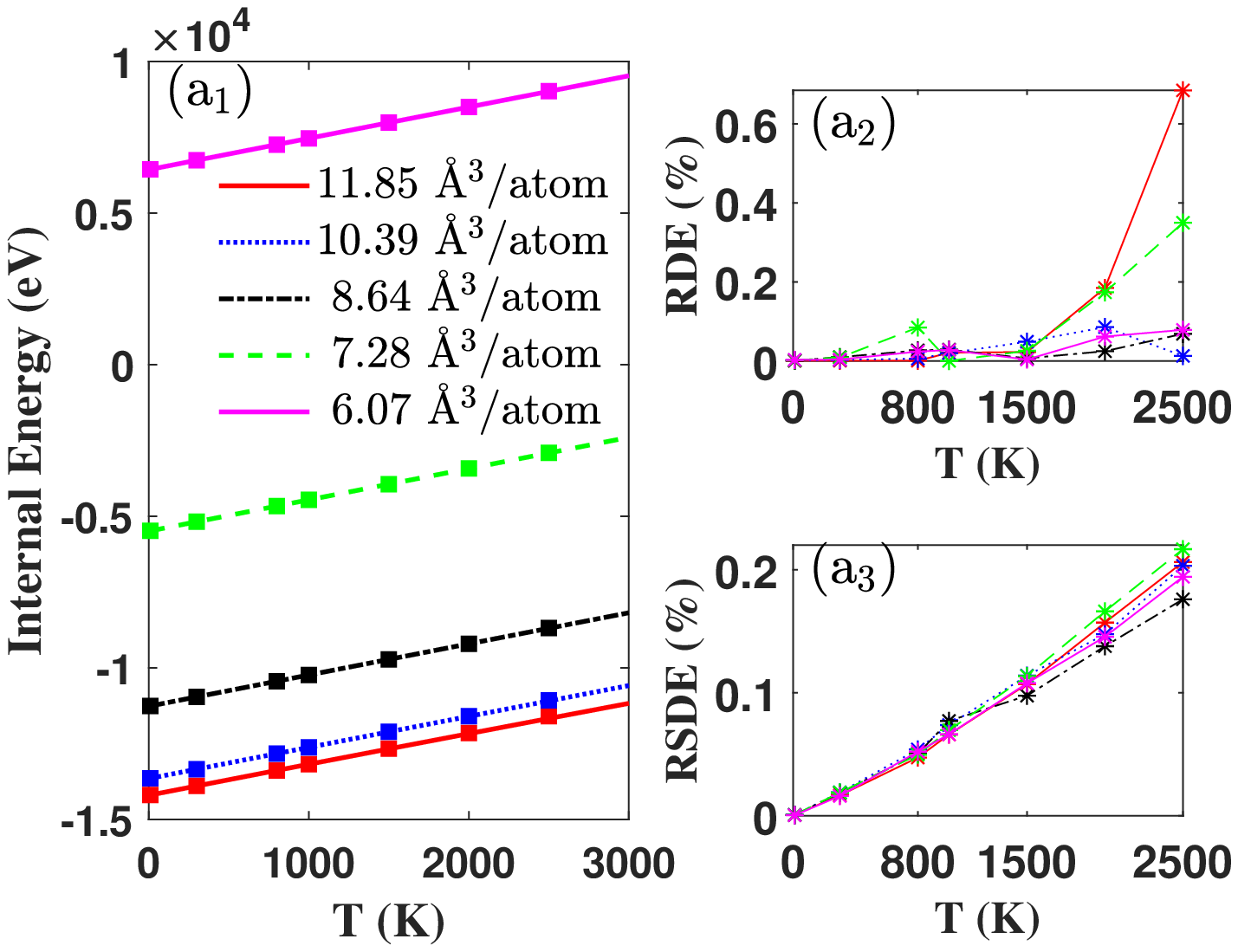}
\label{fig2a}
\centerline{(a)}
\end{minipage}
\begin{minipage}[t]{0.507\linewidth}
\centering
\includegraphics[width=1.68in,height=1.6in]{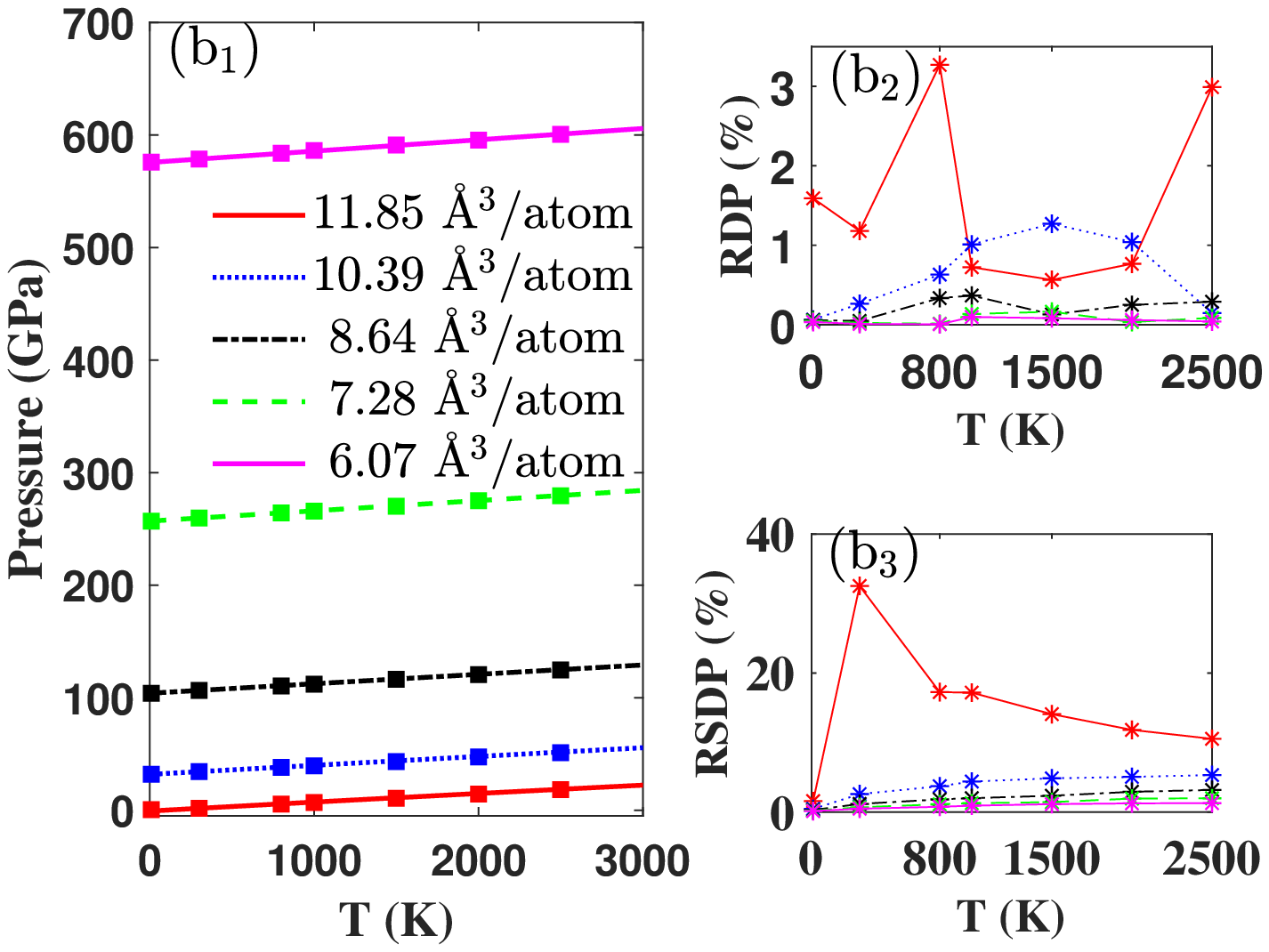}
\label{fig2b}
\centerline{(b)}
\end{minipage}
\caption
{
For solid Cu,
the internal energies (a$_1$) and pressures (b$_1$) obtained from the PF (colored lines) and MD simulations (colored squares).
The relative difference of the internal energy (RDE) and the relative standard deviation (RSDE) of the MD simulations
are shown in (a$_2$) and (a$_3$), respectively,
and the relative difference of the pressure (RDP) and
the relative standard deviations (RSDP) of the MD simulations are shown in (b$_2$) and (b$_3$) respectively.
\label{fig2}
}
\end{figure}

As shown in Fig.\ref{fig2},
for the solid systems with atomic volume ranging from $11.85$ to $6.07$ {\AA}\textsuperscript{3}/atom,
the internal energy $E_{PF}$ and pressure $P_{PF}$ derived from the PF are in excellent agreement
with those ($E_{MD}$ and $P_{MD}$) obtained by MD simulations at temperatures from $10$K to $2500$K.
For $T\leq1000$K,
the relative difference of internal energy RDE$=|(E_{PF}-E_{MD})/E_{MD}|$) is less than $0.09\%$
((Fig.\ref{fig2}a, see data in Supplementary Information).
As the temperature rises up to $2500$K,
the difference gets a bit larger (the maximum is $0.69\%$),
which may be attributed to the statistical fluctuations of MD simulations
because the relative standard deviations of internal energy (RSDE), $\sim0.2\%$,
increase by about three times of those ($\sim0.08\%$) below $1000$K.
As shown in Fig.\ref{fig2}b,
the relative difference of pressure (RDP$=|(P_{PF}-P_{MD})/P_{MD}|$)
is on the level of $\sim2\%$ or less,
which is larger than the RDE.
The reason should be that the MD method for statistic of pressure needs more simulation time
since the relative standard deviations of pressure (RSDP), $\sim10\%$,
are about two orders larger than the RSDE.
In order to confirm this conjecture,
we did similar computations except that the TB potential was replaced with a Lennard-Jones (L-J) potential\cite{LJpotential},
for which the Virial theorem can be applied in the MD simulations to calculate the pressure exactly.
For the volumes of the system ranging from $11.85$ to $7.54$ {\AA}\textsuperscript{3}/atom at $300$K,
the RSDP reduces down to $\sim0.03\%$,
and correspondingly,
the RDP gets smaller to a level of $\sim0.1\%$ (see data in Supplementary Information).
\begin{figure}
\begin{minipage}[t]{0.481\linewidth}
\centering
\includegraphics[width=1.68in,height=1.6in]{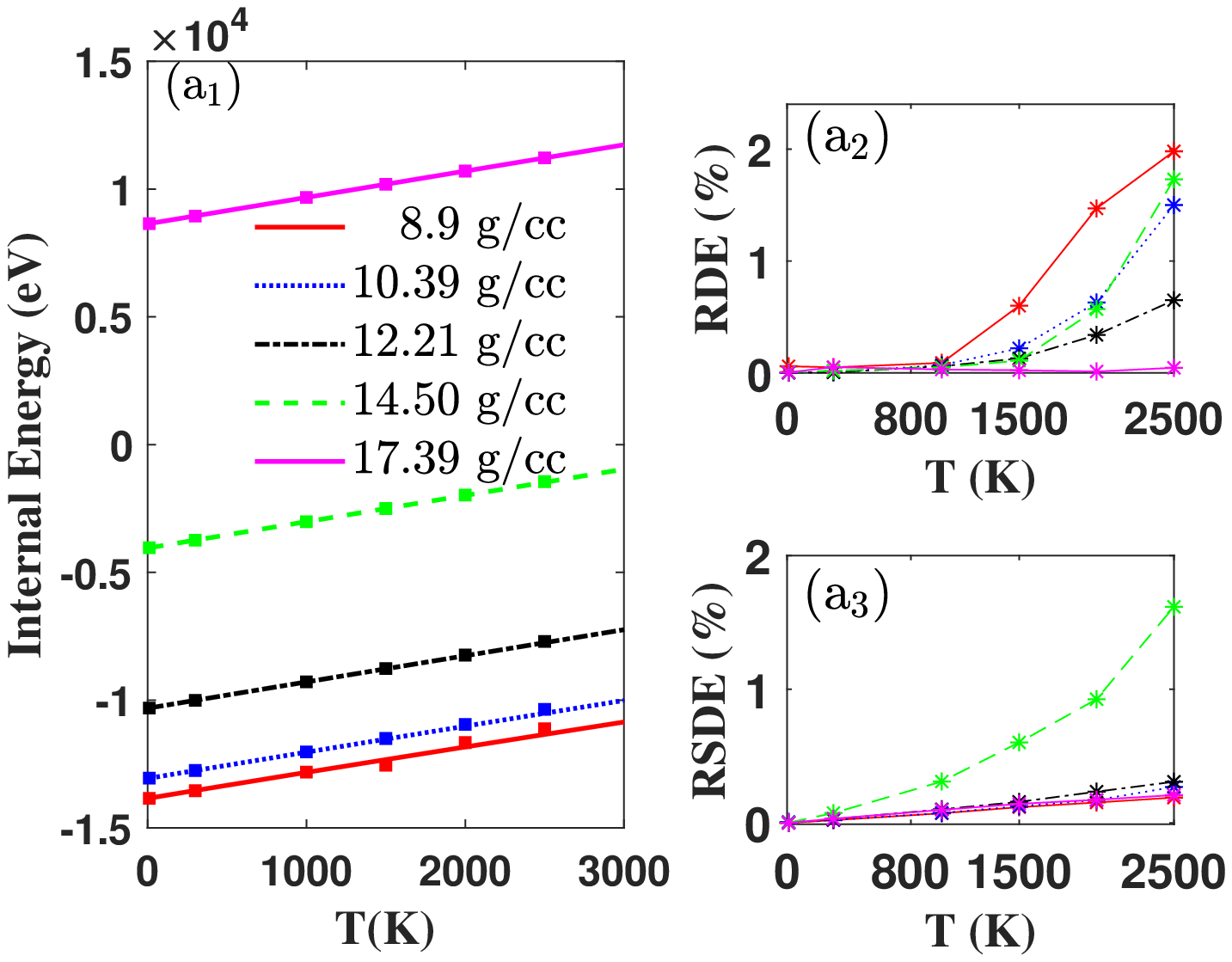}
\label{fig3a}
\centerline{(a)}
\end{minipage}
\begin{minipage}[t]{0.507\linewidth}
\centering
\includegraphics[width=1.68in,height=1.6in]{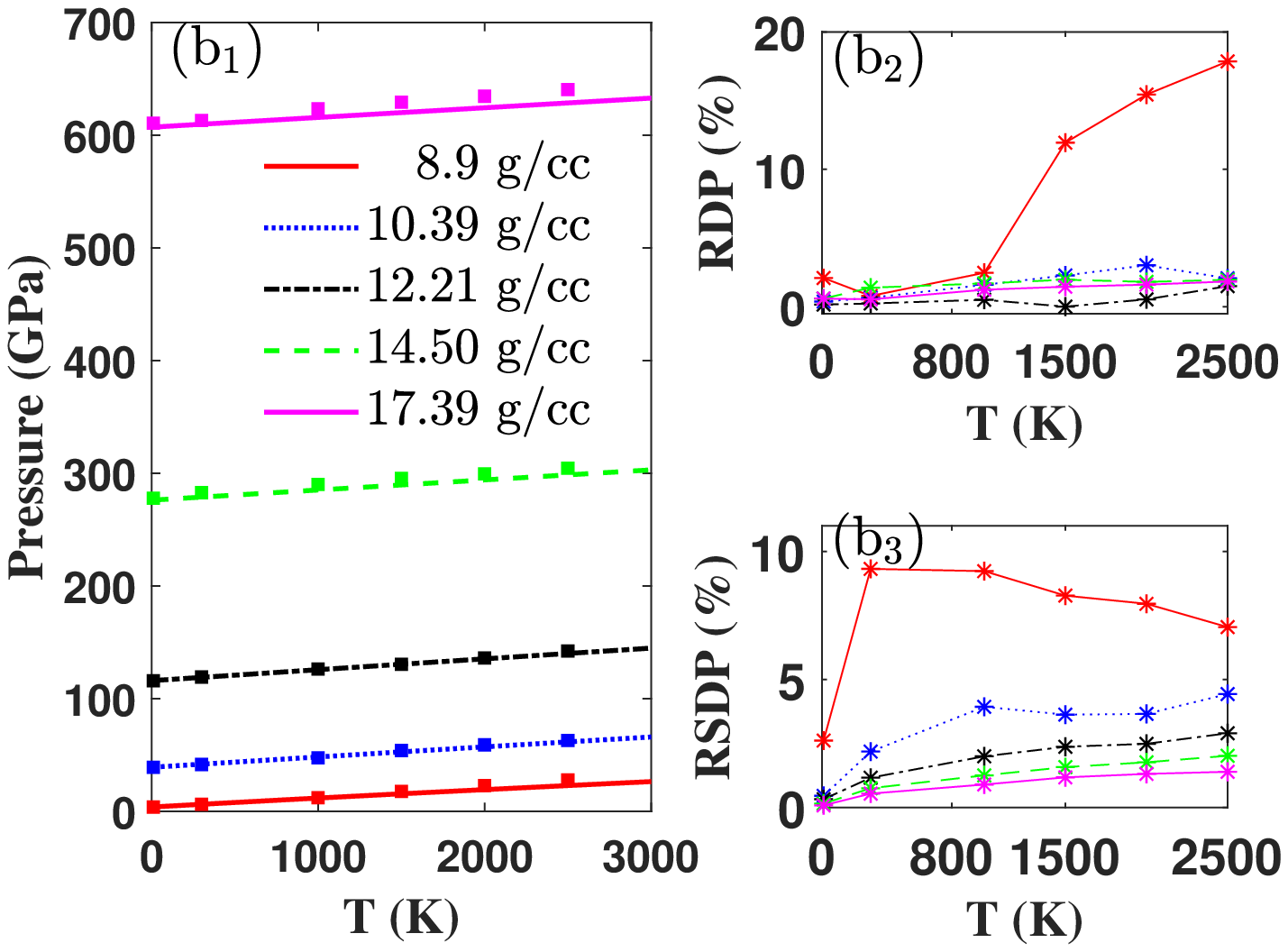}
\label{fig3b}
\centerline{(b)}
\end{minipage}
\caption
{
Similar to Fig.\ref{fig2} except for the liquid state of the systems with different densities.
\label{fig3}
}
\end{figure}

Fig.\ref{fig3} shows that
the internal energy and pressure of liquid Cu obtained by DIA are in good agreement with the MD simulations.
The RDE for all the systems is less than $2\%$,
which is about ten times larger than that for solid Cu.
This difference may stem from the fact that
the $U'(0\ldots q'_i\ldots0)$ felt by a liquid atom differs a little from that felt by other atoms,
which is not the same as the situation in the solid systems.
So, we can perform DIA for more liquid atoms to obtain more accurate results.
As to the pressure,
the RDP is less than $2\%$ for most systems except for the one with a density of $8.9$ g/cc,
which displays larger RDP and can be understood because the RSDP is apparently larger than others (see Fig.\ref{fig3}(b$_3$)).

It should be noted that the precision of DIA is so high that it has reached limit of the MD simulations.
For instance,
the internal energy obtained by DIA of solid Cu with the atomic volume of $11.85${\AA}$^3$/atom
is $-13894.76$eV for $300$K and $-13383.00$eV for $800$K,
which are almost the same as $-13894.96$eV and $-13383.11$eV obtained by the MD simulations
with RSDE of $\sim0.4$\textperthousand.
The gradual increases of RDE and RDP should be attributed to
the fluctuation rise of the MD simulations (see Fig.\ref{fig2}(a$_{3}$) and (b$_{3}$)),
for which smaller time step should be applied to integrate the equations of motion at high temperatures to ensure the precision.

\begin{figure}
\centering
{
\includegraphics[width=2.3in,height=1.8in]{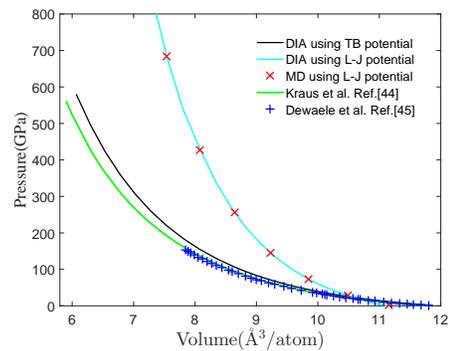}%
}
\caption
{
Isothermal EOSs of Cu at $298$K from the PF using TB\cite{tb} (black solid line)  and L-J\cite{LJpotential} (cyan solid line) potentials for solid Cu are compared to experimental data\cite{copper1,copper2}.
The MD simulations with L-J potential are shown in red crosses.
\label{fig4}
}
\end{figure}
As shown in Fig.\ref{fig4},
the isothermal EOS of solid Cu derived from the PF using TB potential\cite{tb} exhibits the same trend as the experimental results
though the pressures are about $10\%$ larger than the measured values\cite{copper1,copper2}.
This discrepancy should be attributed to the inaccuracy of the empirical potential
because the pressures obtained by DIA and MD simulations coincide with each other quite well.
By contrast, we also used the L-J potential\cite{LJpotential} to calculate the EOS by DIA.
Although the outcome coincides well with the corresponding MD simulations,
it deviates much more from the experiments.
Accordingly,
it calls for a more accurate interatomic potential,
which may resort to \emph{ab initio} calculations in the future.

\subsection{For Solid Argon}
\label{sec.res.ar}
Considering that solid argon (Ar) has been extensively studied experimentally and is believed to be well characterized by L-J pair potential\cite{liquids},
we calculated the PF of solid Ar by DIA to produce the EOS and compared the results with experiments.
To our best knowledge,
no accurate EOS has been put forward for solid Ar by directly solving the PF,
though various EOSs towards Ar systems have been put forward either based on a fitting-parameter procedure (see detailed reviews in Refs.\cite{areos,areos2}),
or in a MC sampling way for disordered liquid and gas states\cite{arwl,nestsample4}.
\begin{figure}
{
\begin{minipage}[t]{0.45\linewidth}
\includegraphics[width=1.6in,height=1.6in]{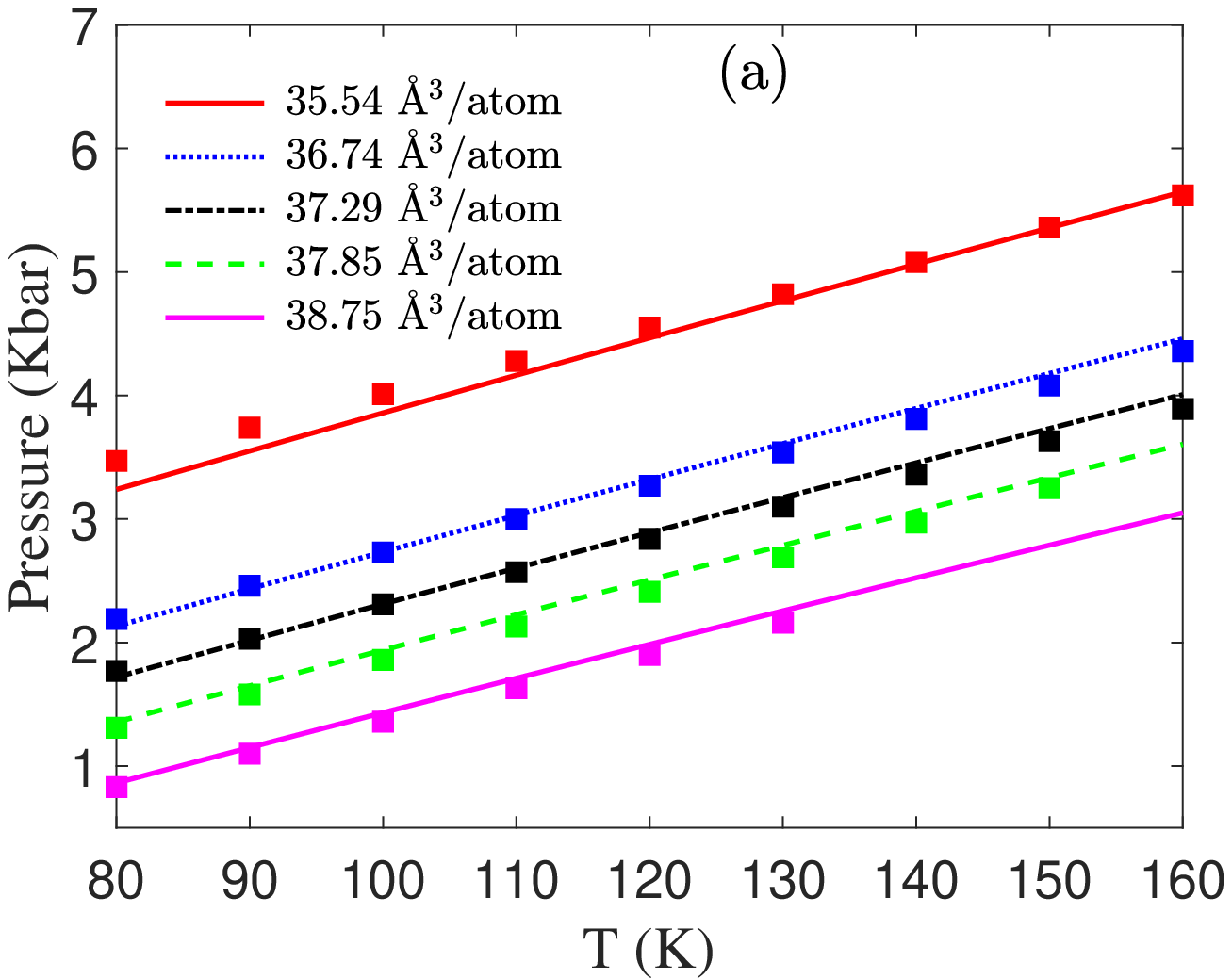}
\label{fig5a}
\end{minipage}
}
{
\begin{minipage}[t]{0.507\linewidth}
\includegraphics[width=1.6in,height=1.6in]{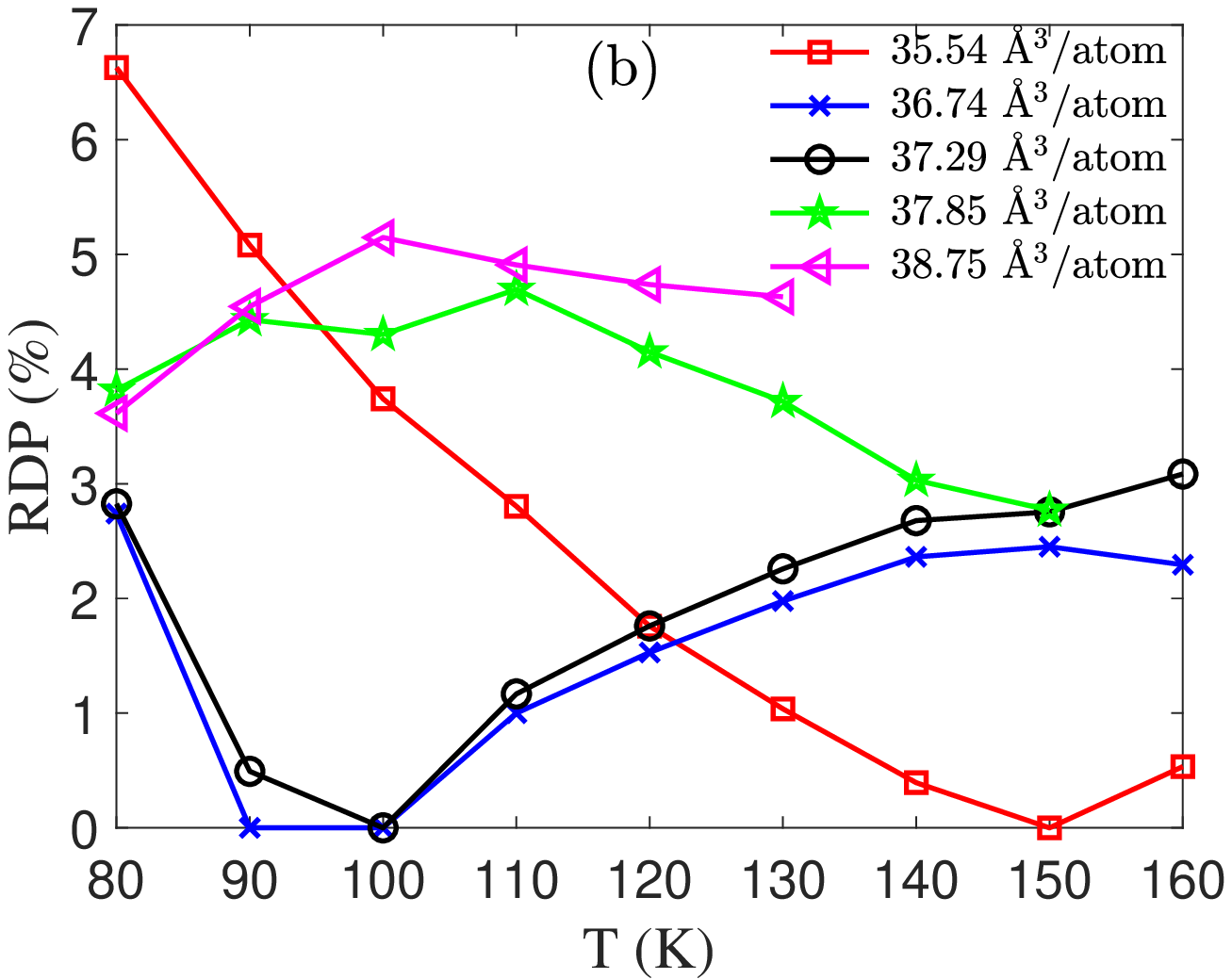}
\label{fig5b}
\end{minipage}
}
\caption{
(a) Isochoric EOSs of solid argon from the PF by DIA (colored lines) and experimental data\cite{isochoric} (colored squares),
and (b) the corresponding relative difference of pressure (RDP).
\label{fig5}
}
\end{figure}

In our work,
the system consists of $32000$ argon atoms placed at FCC sites
and the computational procedure of DIA was the same as described in the system of solid Cu according to Eq.(\ref{eq8})
except for the interatomic potential being replaced with L-J potential\cite{liquids}.
As shown in Fig.\ref{fig5}(a) (see data in Supplementary Information),
the pressure ($P_{PF}$) derived from the PF coincides quite well with the experimental measurements ($P_{EXP}$)\cite{isochoric}.
For the atomic volume ranging from $35.54$ to $38.75$ {\AA}\textsuperscript{3}/atom,
the relative difference (RDP$=|(P_{PF}-P_{EXP})/P_{EXP}|$) between the theoretical pressures
and the experimental ones is within $5\%$ in most conditions as demonstrated in Fig.\ref{fig5}(b).
The maximum RDP of $6.63\%$ occurs in the system with an atomic volume of $V=35.54$ {\AA}\textsuperscript{3}/atom at $80$K,
and may result from the fact,
according to the original reference\cite{isochoric},
that the uncertainty of the measured pressure ranges from $\pm50$ bars above $80$K and becomes larger for lower temperatures.
Considering the possible influence from the empirical L-J potentials,
we may reach a conclusion that the isochoric EOS obtained by DIA is in a good agreement with the experiment.

\subsection{Comparisons with Nested Sampling}
\label{sec.comp}
It is very necessary and interesting to compare DIA with NS,
the state-of-the-art approach to PF,
in terms of computational efficiency and precision.
On the efficiency,
the computational cost of the two methods is determined by the number of times, $N_c$,
to calculate the total potential energy.
In previous work employing NS in solid Al (or NiTi) systems of $64$ atoms\cite{nestsample4},
$N_c$ is about $10^9$ (or $10^{10})$,
while $N_c$ for DIA applied in the solid Cu of $4000$ atoms is $1.5\times10^4$,
showing that DIA is at least four orders faster.
The real computer time of DIA to calculate the PF of $4000$ Cu atoms characterized by the many-body TB potential\cite{tb} 
is about $5$ minutes with full use of a desktop 8-core AMD Ryzen $1800$X CPU ($3.6$GHz per core), 
and, for the $32000$ Ar atoms characterized by the pairwise L-J potential\cite{liquids}, 
is about $2$ minutes using one physical core of the CPU. 
For a solid argon system of $500$ atoms described by L-J potential at temperatures ranging from $80$K to $300$K,
we ran a NS algorithm with $N_c\sim10^9$ and DIA with $N_c\sim10^4$ 
(real computer time is about $4$ hours for NS and about $5$ seconds for DIA respectively by using one physical core of the CPU) 
to calculate internal energy and pressure,
which were compared to the MD simulations using the same potential function, 
demonstrating that the precision of DIA is about $10$ times higher\cite{glcarxiv}.
The accuracy of DIA has also been proved by calculating the internal energy
of graphene or $\gamma$-graphyne materials on Cu substrate using Brener potential function\cite{lyparxiv},
and silicene on Ag substrate using Tersoff potential function\cite{lyparxiv2}.
Due to the ultrahigh efficiency,
DIA has been successfully applied to predict the optimal conditions
for silicene growth on Ag substrate with \emph{ab initio} calculations\cite{lyparxiv2},
which should be the first time
to calculate the absolute free energy at finite temperatures up to thousands Kelvins with first-principle laws.
\section{Conclusion}
\label{sec.con}
In summary,
by our reinterpretation of integral,
DIA to PF of large molecules and condensed systems was established.
The accuracy of DIA was strictly validated by vast MD simulations
for C$_{60}$ clusters, condensed Cu, solid argon, graphene and silicene on substrate,
and by experiments for solid Ar respectively.
Compared to state-of-the-art method for PF,
DIA works at least four orders faster with about one order more precise.
The new approach will find its vast applications
in investigating the products of complex chemical reactions, 
thermodynamic properties of large molecules and macroscopic systems,
which highly relates to designing novel material,
predicting various phase transitions
and parameter-free EOS under extreme conditions.

\section{acknowledgement}
TCW acknowledges the support from National Natural Science Foundation of China under Grant No.21727801.


\begin{thebibliography}{60}%
\makeatletter
\providecommand \@ifxundefined [1]{%
 \@ifx{#1\undefined}
}%
\providecommand \@ifnum [1]{%
 \ifnum #1\expandafter \@firstoftwo
 \else \expandafter \@secondoftwo
 \fi
}%
\providecommand \@ifx [1]{%
 \ifx #1\expandafter \@firstoftwo
 \else \expandafter \@secondoftwo
 \fi
}%
\providecommand \natexlab [1]{#1}%
\providecommand \enquote  [1]{``#1''}%
\providecommand \bibnamefont  [1]{#1}%
\providecommand \bibfnamefont [1]{#1}%
\providecommand \citenamefont [1]{#1}%
\providecommand \href@noop [0]{\@secondoftwo}%
\providecommand \href [0]{\begingroup \@sanitize@url \@href}%
\providecommand \@href[1]{\@@startlink{#1}\@@href}%
\providecommand \@@href[1]{\endgroup#1\@@endlink}%
\providecommand \@sanitize@url [0]{\catcode `\\12\catcode `\$12\catcode
  `\&12\catcode `\#12\catcode `\^12\catcode `\_12\catcode `\%12\relax}%
\providecommand \@@startlink[1]{}%
\providecommand \@@endlink[0]{}%
\providecommand \url  [0]{\begingroup\@sanitize@url \@url }%
\providecommand \@url [1]{\endgroup\@href {#1}{\urlprefix }}%
\providecommand \urlprefix  [0]{URL }%
\providecommand \Eprint [0]{\href }%
\providecommand \doibase [0]{http://dx.doi.org/}%
\providecommand \selectlanguage [0]{\@gobble}%
\providecommand \bibinfo  [0]{\@secondoftwo}%
\providecommand \bibfield  [0]{\@secondoftwo}%
\providecommand \translation [1]{[#1]}%
\providecommand \BibitemOpen [0]{}%
\providecommand \bibitemStop [0]{}%
\providecommand \bibitemNoStop [0]{.\EOS\space}%
\providecommand \EOS [0]{\spacefactor3000\relax}%
\providecommand \BibitemShut  [1]{\csname bibitem#1\endcsname}%
\let\auto@bib@innerbib\@empty
\bibitem [{\citenamefont {Chipot}\ and\ \citenamefont {Pohorille}(2007)}]{fe1}%
  \BibitemOpen
  \bibinfo {editor} {\bibfnamefont {C.}~\bibnamefont {Chipot}}\ and\ \bibinfo
  {editor} {\bibfnamefont {A.}~\bibnamefont {Pohorille}},\ eds.,\ \href@noop {}
  {\emph {\bibinfo {title} {Free Energy Calculations: Theory and Applications
  in Chmistry and Biology}}}\ (\bibinfo  {publisher} {Springer Berlin},\
  \bibinfo {year} {2007})\BibitemShut {NoStop}%
\bibitem [{\citenamefont {D~Stacey}(2005)}]{eos1}%
  \BibitemOpen
  \bibfield  {author} {\bibinfo {author} {\bibfnamefont {F.}~\bibnamefont
  {D~Stacey}},\ }\href@noop {} {\bibfield  {journal} {\bibinfo  {journal} {Rep.
  Prog. Phys.}\ }\textbf {\bibinfo {volume} {68}},\ \bibinfo {pages} {341}
  (\bibinfo {year} {2005})}\BibitemShut {NoStop}%
\bibitem [{\citenamefont {Ross}\ and\ \citenamefont {Young}(1993)}]{eos2}%
  \BibitemOpen
  \bibfield  {author} {\bibinfo {author} {\bibfnamefont {M.}~\bibnamefont
  {Ross}}\ and\ \bibinfo {author} {\bibfnamefont {D.~A.}\ \bibnamefont
  {Young}},\ }\href {\doibase 10.1146/annurev.pc.44.100193.000425} {\bibfield
  {journal} {\bibinfo  {journal} {Annu. Rev. Phys. Chem.}\ }\textbf {\bibinfo
  {volume} {44}},\ \bibinfo {pages} {61} (\bibinfo {year} {1993})}\BibitemShut
  {NoStop}%
\bibitem [{\citenamefont {Monson}\ and\ \citenamefont {Kofke}(2000)}]{kofke}%
  \BibitemOpen
  \bibfield  {author} {\bibinfo {author} {\bibfnamefont {P.~A.}\ \bibnamefont
  {Monson}}\ and\ \bibinfo {author} {\bibfnamefont {D.~A.}\ \bibnamefont
  {Kofke}},\ }\href@noop {} {\bibfield  {journal} {\bibinfo  {journal} {Adv.
  Chem. Phys.}\ }\textbf {\bibinfo {volume} {115}},\ \bibinfo {pages} {113}
  (\bibinfo {year} {2000})}\BibitemShut {NoStop}%
\bibitem [{\citenamefont {Ushcats}\ \emph {et~al.}(2016)\citenamefont
  {Ushcats}, \citenamefont {Bulavin}, \citenamefont {Sysoev}, \citenamefont
  {Bardik},\ and\ \citenamefont {Alekseev}}]{pfconden}%
  \BibitemOpen
  \bibfield  {author} {\bibinfo {author} {\bibfnamefont {M.~V.}\ \bibnamefont
  {Ushcats}}, \bibinfo {author} {\bibfnamefont {L.~A.}\ \bibnamefont
  {Bulavin}}, \bibinfo {author} {\bibfnamefont {V.~M.}\ \bibnamefont {Sysoev}},
  \bibinfo {author} {\bibfnamefont {V.~Y.}\ \bibnamefont {Bardik}}, \ and\
  \bibinfo {author} {\bibfnamefont {A.~N.}\ \bibnamefont {Alekseev}},\
  }\href@noop {} {\bibfield  {journal} {\bibinfo  {journal} {J. Mol. Liq.}\
  }\textbf {\bibinfo {volume} {224}},\ \bibinfo {pages} {694 } (\bibinfo {year}
  {2016})}\BibitemShut {NoStop}%
\bibitem [{\citenamefont {Allen}\ and\ \citenamefont
  {Tildesley}(1987)}]{liquids}%
  \BibitemOpen
  \bibfield  {author} {\bibinfo {author} {\bibfnamefont {M.~P.}\ \bibnamefont
  {Allen}}\ and\ \bibinfo {author} {\bibfnamefont {D.~J.}\ \bibnamefont
  {Tildesley}},\ }\href@noop {} {\emph {\bibinfo {title} {Computer Simulation
  of Liquids}}}\ (\bibinfo  {publisher} {Oxford University Press},\ \bibinfo
  {year} {1987})\BibitemShut {NoStop}%
\bibitem [{\citenamefont {Rapaort}(2004)}]{MD}%
  \BibitemOpen
  \bibfield  {author} {\bibinfo {author} {\bibfnamefont {D.~C.}\ \bibnamefont
  {Rapaort}},\ }\href@noop {} {\emph {\bibinfo {title} {The Art of Molecular
  Dynamics Simulation}}}\ (\bibinfo  {publisher} {Cambridge University Press},\
  \bibinfo {year} {2004})\BibitemShut {NoStop}%
\bibitem [{\citenamefont {Landau}\ and\ \citenamefont {Kurt}(2005)}]{MC}%
  \BibitemOpen
  \bibfield  {author} {\bibinfo {author} {\bibfnamefont {D.~P.}\ \bibnamefont
  {Landau}}\ and\ \bibinfo {author} {\bibfnamefont {B.}~\bibnamefont {Kurt}},\
  }\href@noop {} {\emph {\bibinfo {title} {A Guide to Monte Carlo Simulations
  in Statistical Physics}}}\ (\bibinfo  {publisher} {Cambridge University
  Press},\ \bibinfo {year} {2005})\BibitemShut {NoStop}%
\bibitem [{\citenamefont {Allen}\ and\ \citenamefont {Frenkel}(1989)}]{mcmd}%
  \BibitemOpen
  \bibfield  {author} {\bibinfo {author} {\bibfnamefont {M.~P.}\ \bibnamefont
  {Allen}}\ and\ \bibinfo {author} {\bibfnamefont {D.}~\bibnamefont
  {Frenkel}},\ }\href@noop {} {\bibfield  {journal} {\bibinfo  {journal}
  {Comput. Phys. Rep.}\ }\textbf {\bibinfo {volume} {9}},\ \bibinfo {pages}
  {301} (\bibinfo {year} {1989})}\BibitemShut {NoStop}%
\bibitem [{\citenamefont {Ballard}\ \emph {et~al.}(2015)\citenamefont
  {Ballard}, \citenamefont {Martiniani}, \citenamefont {Stevenson},
  \citenamefont {Somani},\ and\ \citenamefont {Wales}}]{pes1}%
  \BibitemOpen
  \bibfield  {author} {\bibinfo {author} {\bibfnamefont {A.~J.}\ \bibnamefont
  {Ballard}}, \bibinfo {author} {\bibfnamefont {S.}~\bibnamefont {Martiniani}},
  \bibinfo {author} {\bibfnamefont {J.~D.}\ \bibnamefont {Stevenson}}, \bibinfo
  {author} {\bibfnamefont {S.}~\bibnamefont {Somani}}, \ and\ \bibinfo {author}
  {\bibfnamefont {D.~J.}\ \bibnamefont {Wales}},\ }\href {\doibase
  10.1002/wcms.1217} {\bibfield  {journal} {\bibinfo  {journal} {WIREs. Comput.
  Mol. Sci.}\ }\textbf {\bibinfo {volume} {5}},\ \bibinfo {pages} {273}
  (\bibinfo {year} {2015})}\BibitemShut {NoStop}%
\bibitem [{\citenamefont {Jarzynski}(1997)}]{jarz1997}%
  \BibitemOpen
  \bibfield  {author} {\bibinfo {author} {\bibfnamefont {C.}~\bibnamefont
  {Jarzynski}},\ }\href {\doibase 10.1103/PhysRevLett.78.2690} {\bibfield
  {journal} {\bibinfo  {journal} {Phys. Rev. Lett.}\ }\textbf {\bibinfo
  {volume} {78}},\ \bibinfo {pages} {2690} (\bibinfo {year}
  {1997})}\BibitemShut {NoStop}%
\bibitem [{\citenamefont {Moustafa}\ \emph {et~al.}(2015)\citenamefont
  {Moustafa}, \citenamefont {Schultz},\ and\ \citenamefont {Kofke}}]{kofke15}%
  \BibitemOpen
  \bibfield  {author} {\bibinfo {author} {\bibfnamefont {S.~G.}\ \bibnamefont
  {Moustafa}}, \bibinfo {author} {\bibfnamefont {A.~J.}\ \bibnamefont
  {Schultz}}, \ and\ \bibinfo {author} {\bibfnamefont {D.~A.}\ \bibnamefont
  {Kofke}},\ }\href {\doibase 10.1103/PhysRevE.92.043303} {\bibfield  {journal}
  {\bibinfo  {journal} {Phys. Rev. E}\ }\textbf {\bibinfo {volume} {92}},\
  \bibinfo {pages} {043303} (\bibinfo {year} {2015})}\BibitemShut {NoStop}%
\bibitem [{\citenamefont {Wang}\ and\ \citenamefont {Landau}(2001)}]{wl1}%
  \BibitemOpen
  \bibfield  {author} {\bibinfo {author} {\bibfnamefont {F.}~\bibnamefont
  {Wang}}\ and\ \bibinfo {author} {\bibfnamefont {D.~P.}\ \bibnamefont
  {Landau}},\ }\href {\doibase 10.1103/PhysRevLett.86.2050} {\bibfield
  {journal} {\bibinfo  {journal} {Phys. Rev. Lett.}\ }\textbf {\bibinfo
  {volume} {86}},\ \bibinfo {pages} {2050} (\bibinfo {year}
  {2001})}\BibitemShut {NoStop}%
\bibitem [{\citenamefont {Singh}\ \emph {et~al.}(2012)\citenamefont {Singh},
  \citenamefont {Chopra},\ and\ \citenamefont {de~Pablo}}]{DOS}%
  \BibitemOpen
  \bibfield  {author} {\bibinfo {author} {\bibfnamefont {S.}~\bibnamefont
  {Singh}}, \bibinfo {author} {\bibfnamefont {M.}~\bibnamefont {Chopra}}, \
  and\ \bibinfo {author} {\bibfnamefont {J.~J.}\ \bibnamefont {de~Pablo}},\
  }\href {\doibase 10.1146/annurev-chembioeng-062011-081032} {\bibfield
  {journal} {\bibinfo  {journal} {Annu. Rev. Chem. Biomol. Eng.}\ }\textbf
  {\bibinfo {volume} {3}},\ \bibinfo {pages} {369} (\bibinfo {year}
  {2012})}\BibitemShut {NoStop}%
\bibitem [{\citenamefont {Li}\ \emph {et~al.}(2016)\citenamefont {Li},
  \citenamefont {Ning}, \citenamefont {Zhuang},\ and\ \citenamefont
  {Ning}}]{nxj}%
  \BibitemOpen
  \bibfield  {author} {\bibinfo {author} {\bibfnamefont {J.-T.}\ \bibnamefont
  {Li}}, \bibinfo {author} {\bibfnamefont {B.-Y.}\ \bibnamefont {Ning}},
  \bibinfo {author} {\bibfnamefont {J.}~\bibnamefont {Zhuang}}, \ and\ \bibinfo
  {author} {\bibfnamefont {X.-J.}\ \bibnamefont {Ning}},\ }\href {\doibase
  10.1088/1674-1056/26/3/030501} {\bibfield  {journal} {\bibinfo  {journal}
  {Chin. Phys. B}\ }\textbf {\bibinfo {volume} {26}},\ \bibinfo {pages}
  {030501} (\bibinfo {year} {2016})}\BibitemShut {NoStop}%
\bibitem [{\citenamefont {Berg}\ and\ \citenamefont
  {Neuhaus}(1991)}]{multicanonical}%
  \BibitemOpen
  \bibfield  {author} {\bibinfo {author} {\bibfnamefont {B.~A.}\ \bibnamefont
  {Berg}}\ and\ \bibinfo {author} {\bibfnamefont {T.}~\bibnamefont {Neuhaus}},\
  }\href@noop {} {\bibfield  {journal} {\bibinfo  {journal} {Phys. Lett. B}\
  }\textbf {\bibinfo {volume} {267}},\ \bibinfo {pages} {249} (\bibinfo {year}
  {1991})}\BibitemShut {NoStop}%
\bibitem [{\citenamefont {Wang}\ \emph {et~al.}(1999)\citenamefont {Wang},
  \citenamefont {Tay},\ and\ \citenamefont {Swendsen}}]{transitionmatrix}%
  \BibitemOpen
  \bibfield  {author} {\bibinfo {author} {\bibfnamefont {J.-S.}\ \bibnamefont
  {Wang}}, \bibinfo {author} {\bibfnamefont {T.~K.}\ \bibnamefont {Tay}}, \
  and\ \bibinfo {author} {\bibfnamefont {R.~H.}\ \bibnamefont {Swendsen}},\
  }\href {\doibase 10.1103/PhysRevLett.82.476} {\bibfield  {journal} {\bibinfo
  {journal} {Phys. Rev. Lett.}\ }\textbf {\bibinfo {volume} {82}},\ \bibinfo
  {pages} {476} (\bibinfo {year} {1999})}\BibitemShut {NoStop}%
\bibitem [{\citenamefont {Skilling}(2004)}]{nestsample1}%
  \BibitemOpen
  \bibfield  {author} {\bibinfo {author} {\bibfnamefont {J.}~\bibnamefont
  {Skilling}},\ }\href@noop {} {\bibfield  {journal} {\bibinfo  {journal} {AIP
  Conf. Proc.}\ }\textbf {\bibinfo {volume} {735}},\ \bibinfo {pages} {395}
  (\bibinfo {year} {2004})}\BibitemShut {NoStop}%
\bibitem [{\citenamefont {P\'{a}rtay}\ \emph {et~al.}(2010)\citenamefont
  {P\'{a}rtay}, \citenamefont {Bart\'{o}k},\ and\ \citenamefont
  {Cs\'{a}nyi}}]{nestsample3}%
  \BibitemOpen
  \bibfield  {author} {\bibinfo {author} {\bibfnamefont {L.~B.}\ \bibnamefont
  {P\'{a}rtay}}, \bibinfo {author} {\bibfnamefont {A.~P.}\ \bibnamefont
  {Bart\'{o}k}}, \ and\ \bibinfo {author} {\bibfnamefont {G.}~\bibnamefont
  {Cs\'{a}nyi}},\ }\href {\doibase 10.1021/jp1012973} {\bibfield  {journal}
  {\bibinfo  {journal} {J. Phys. Chem. B}\ }\textbf {\bibinfo {volume} {114}},\
  \bibinfo {pages} {10502} (\bibinfo {year} {2010})}\BibitemShut {NoStop}%
\bibitem [{\citenamefont {Nielsen}(2013)}]{nestsample7}%
  \BibitemOpen
  \bibfield  {author} {\bibinfo {author} {\bibfnamefont {S.~O.}\ \bibnamefont
  {Nielsen}},\ }\href@noop {} {\bibfield  {journal} {\bibinfo  {journal} {J.
  Chem. Phys.}\ }\textbf {\bibinfo {volume} {139}},\ \bibinfo {pages} {124104}
  (\bibinfo {year} {2013})}\BibitemShut {NoStop}%
\bibitem [{\citenamefont {Wilson}\ \emph {et~al.}(2015)\citenamefont {Wilson},
  \citenamefont {Gelb},\ and\ \citenamefont {Nielsen}}]{nestsample5}%
  \BibitemOpen
  \bibfield  {author} {\bibinfo {author} {\bibfnamefont {B.~A.}\ \bibnamefont
  {Wilson}}, \bibinfo {author} {\bibfnamefont {L.~D.}\ \bibnamefont {Gelb}}, \
  and\ \bibinfo {author} {\bibfnamefont {S.~O.}\ \bibnamefont {Nielsen}},\
  }\href@noop {} {\bibfield  {journal} {\bibinfo  {journal} {J. Chem. Phys.}\
  }\textbf {\bibinfo {volume} {143}},\ \bibinfo {pages} {154108} (\bibinfo
  {year} {2015})}\BibitemShut {NoStop}%
\bibitem [{\citenamefont {P\'artay}\ \emph {et~al.}(2014)\citenamefont
  {P\'artay}, \citenamefont {Bart\'ok},\ and\ \citenamefont
  {Cs\'anyi}}]{nestsample10}%
  \BibitemOpen
  \bibfield  {author} {\bibinfo {author} {\bibfnamefont {L.~B.}\ \bibnamefont
  {P\'artay}}, \bibinfo {author} {\bibfnamefont {A.~P.}\ \bibnamefont
  {Bart\'ok}}, \ and\ \bibinfo {author} {\bibfnamefont {G.}~\bibnamefont
  {Cs\'anyi}},\ }\href {\doibase 10.1103/PhysRevE.89.022302} {\bibfield
  {journal} {\bibinfo  {journal} {Phys. Rev. E}\ }\textbf {\bibinfo {volume}
  {89}},\ \bibinfo {pages} {022302} (\bibinfo {year} {2014})}\BibitemShut
  {NoStop}%
\bibitem [{\citenamefont {Do}\ \emph {et~al.}(2011)\citenamefont {Do},
  \citenamefont {Hirst},\ and\ \citenamefont {Wheatley}}]{do1}%
  \BibitemOpen
  \bibfield  {author} {\bibinfo {author} {\bibfnamefont {H.}~\bibnamefont
  {Do}}, \bibinfo {author} {\bibfnamefont {J.~D.}\ \bibnamefont {Hirst}}, \
  and\ \bibinfo {author} {\bibfnamefont {R.~J.}\ \bibnamefont {Wheatley}},\
  }\href@noop {} {\bibfield  {journal} {\bibinfo  {journal} {J. Chem. Phys.}\
  }\textbf {\bibinfo {volume} {135}},\ \bibinfo {pages} {174105} (\bibinfo
  {year} {2011})}\BibitemShut {NoStop}%
\bibitem [{\citenamefont {Coe}\ \emph {et~al.}(2009)\citenamefont {Coe},
  \citenamefont {Sewell},\ and\ \citenamefont {Shaw}}]{nestsample9}%
  \BibitemOpen
  \bibfield  {author} {\bibinfo {author} {\bibfnamefont {J.~D.}\ \bibnamefont
  {Coe}}, \bibinfo {author} {\bibfnamefont {T.~D.}\ \bibnamefont {Sewell}}, \
  and\ \bibinfo {author} {\bibfnamefont {M.~S.}\ \bibnamefont {Shaw}},\
  }\href@noop {} {\bibfield  {journal} {\bibinfo  {journal} {J. Chem. Phys.}\
  }\textbf {\bibinfo {volume} {131}},\ \bibinfo {pages} {074105} (\bibinfo
  {year} {2009})}\BibitemShut {NoStop}%
\bibitem [{\citenamefont {Baldock}\ \emph {et~al.}(2016)\citenamefont
  {Baldock}, \citenamefont {P\'artay}, \citenamefont {Bart\'ok}, \citenamefont
  {Payne},\ and\ \citenamefont {Cs\'anyi}}]{nestsample4}%
  \BibitemOpen
  \bibfield  {author} {\bibinfo {author} {\bibfnamefont {R.~J.~N.}\
  \bibnamefont {Baldock}}, \bibinfo {author} {\bibfnamefont {L.~B.}\
  \bibnamefont {P\'artay}}, \bibinfo {author} {\bibfnamefont {A.~P.}\
  \bibnamefont {Bart\'ok}}, \bibinfo {author} {\bibfnamefont {M.~C.}\
  \bibnamefont {Payne}}, \ and\ \bibinfo {author} {\bibfnamefont
  {G.}~\bibnamefont {Cs\'anyi}},\ }\href {\doibase 10.1103/PhysRevB.93.174108}
  {\bibfield  {journal} {\bibinfo  {journal} {Phys. Rev. B}\ }\textbf {\bibinfo
  {volume} {93}},\ \bibinfo {pages} {174108} (\bibinfo {year}
  {2016})}\BibitemShut {NoStop}%
\bibitem [{\citenamefont {Do}\ and\ \citenamefont {Wheatley}(2013)}]{do3}%
  \BibitemOpen
  \bibfield  {author} {\bibinfo {author} {\bibfnamefont {H.}~\bibnamefont
  {Do}}\ and\ \bibinfo {author} {\bibfnamefont {R.~J.}\ \bibnamefont
  {Wheatley}},\ }\href@noop {} {\bibfield  {journal} {\bibinfo  {journal} {J.
  Chem. Theory Comput.}\ }\textbf {\bibinfo {volume} {9}},\ \bibinfo {pages}
  {165} (\bibinfo {year} {2013})}\BibitemShut {NoStop}%
\bibitem [{\citenamefont {Do}\ and\ \citenamefont {Wheatley}(2016)}]{do4}%
  \BibitemOpen
  \bibfield  {author} {\bibinfo {author} {\bibfnamefont {H.}~\bibnamefont
  {Do}}\ and\ \bibinfo {author} {\bibfnamefont {R.~J.}\ \bibnamefont
  {Wheatley}},\ }\href@noop {} {\bibfield  {journal} {\bibinfo  {journal} {J.
  Chem. Phys.}\ }\textbf {\bibinfo {volume} {145}},\ \bibinfo {pages} {084116}
  (\bibinfo {year} {2016})}\BibitemShut {NoStop}%
\bibitem [{\citenamefont {Do}\ \emph {et~al.}(2012)\citenamefont {Do},
  \citenamefont {Hirst},\ and\ \citenamefont {Wheatley}}]{do2}%
  \BibitemOpen
  \bibfield  {author} {\bibinfo {author} {\bibfnamefont {H.}~\bibnamefont
  {Do}}, \bibinfo {author} {\bibfnamefont {J.~D.}\ \bibnamefont {Hirst}}, \
  and\ \bibinfo {author} {\bibfnamefont {R.~J.}\ \bibnamefont {Wheatley}},\
  }\href@noop {} {\bibfield  {journal} {\bibinfo  {journal} {J. Phys. Chem. B}\
  }\textbf {\bibinfo {volume} {116}},\ \bibinfo {pages} {4535} (\bibinfo {year}
  {2012})}\BibitemShut {NoStop}%
\bibitem [{\citenamefont {Burkoff}\ \emph {et~al.}(2012)\citenamefont
  {Burkoff}, \citenamefont {V\'{a}rnai}, \citenamefont {Wells},\ and\
  \citenamefont {Wild}}]{nestsample11}%
  \BibitemOpen
  \bibfield  {author} {\bibinfo {author} {\bibfnamefont {N.~S.}\ \bibnamefont
  {Burkoff}}, \bibinfo {author} {\bibfnamefont {C.}~\bibnamefont {V\'{a}rnai}},
  \bibinfo {author} {\bibfnamefont {S.~A.}\ \bibnamefont {Wells}}, \ and\
  \bibinfo {author} {\bibfnamefont {D.~L.}\ \bibnamefont {Wild}},\ }\href@noop
  {} {\bibfield  {journal} {\bibinfo  {journal} {Biophys. J.}\ }\textbf
  {\bibinfo {volume} {102}},\ \bibinfo {pages} {878} (\bibinfo {year}
  {2012})}\BibitemShut {NoStop}%
\bibitem [{\citenamefont {Baldock}\ \emph {et~al.}(2017)\citenamefont
  {Baldock}, \citenamefont {Bernstein}, \citenamefont {Salerno}, \citenamefont
  {P\'artay},\ and\ \citenamefont {Cs\'anyi}}]{nestsample12}%
  \BibitemOpen
  \bibfield  {author} {\bibinfo {author} {\bibfnamefont {R.~J.~N.}\
  \bibnamefont {Baldock}}, \bibinfo {author} {\bibfnamefont {N.}~\bibnamefont
  {Bernstein}}, \bibinfo {author} {\bibfnamefont {K.~M.}\ \bibnamefont
  {Salerno}}, \bibinfo {author} {\bibfnamefont {L.~B.}\ \bibnamefont
  {P\'artay}}, \ and\ \bibinfo {author} {\bibfnamefont {G.}~\bibnamefont
  {Cs\'anyi}},\ }\href {\doibase 10.1103/PhysRevE.96.043311} {\bibfield
  {journal} {\bibinfo  {journal} {Phys. Rev. E}\ }\textbf {\bibinfo {volume}
  {96}},\ \bibinfo {pages} {043311} (\bibinfo {year} {2017})}\BibitemShut
  {NoStop}%
\bibitem [{\citenamefont {Bolhuis}\ and\ \citenamefont
  {Cs\'anyi}(2018)}]{nestsample13}%
  \BibitemOpen
  \bibfield  {author} {\bibinfo {author} {\bibfnamefont {P.~G.}\ \bibnamefont
  {Bolhuis}}\ and\ \bibinfo {author} {\bibfnamefont {G.}~\bibnamefont
  {Cs\'anyi}},\ }\href {\doibase 10.1103/PhysRevLett.120.250601} {\bibfield
  {journal} {\bibinfo  {journal} {Phys. Rev. Lett.}\ }\textbf {\bibinfo
  {volume} {120}},\ \bibinfo {pages} {250601} (\bibinfo {year}
  {2018})}\BibitemShut {NoStop}%
\bibitem [{\citenamefont {Oganov}\ and\ \citenamefont {Glass}(2006)}]{glopt1}%
  \BibitemOpen
  \bibfield  {author} {\bibinfo {author} {\bibfnamefont {A.~R.}\ \bibnamefont
  {Oganov}}\ and\ \bibinfo {author} {\bibfnamefont {C.~W.}\ \bibnamefont
  {Glass}},\ }\href {\doibase 10.1063/1.2210932} {\bibfield  {journal}
  {\bibinfo  {journal} {J. Chem. Phys.}\ }\textbf {\bibinfo {volume} {124}},\
  \bibinfo {pages} {244704} (\bibinfo {year} {2006})}\BibitemShut {NoStop}%
\bibitem [{\citenamefont {Wales}\ and\ \citenamefont
  {Scheraga}(1999)}]{glopt2}%
  \BibitemOpen
  \bibfield  {author} {\bibinfo {author} {\bibfnamefont {D.~J.}\ \bibnamefont
  {Wales}}\ and\ \bibinfo {author} {\bibfnamefont {H.~A.}\ \bibnamefont
  {Scheraga}},\ }\href {\doibase 10.1126/science.285.5432.1368} {\bibfield
  {journal} {\bibinfo  {journal} {Science}\ }\textbf {\bibinfo {volume}
  {285}},\ \bibinfo {pages} {1368} (\bibinfo {year} {1999})}\BibitemShut
  {NoStop}%
\bibitem [{\citenamefont {Deaven}\ and\ \citenamefont {Ho}(1995)}]{glopt3}%
  \BibitemOpen
  \bibfield  {author} {\bibinfo {author} {\bibfnamefont {D.~M.}\ \bibnamefont
  {Deaven}}\ and\ \bibinfo {author} {\bibfnamefont {K.~M.}\ \bibnamefont
  {Ho}},\ }\href {\doibase 10.1103/PhysRevLett.75.288} {\bibfield  {journal}
  {\bibinfo  {journal} {Phys. Rev. Lett.}\ }\textbf {\bibinfo {volume} {75}},\
  \bibinfo {pages} {288} (\bibinfo {year} {1995})}\BibitemShut {NoStop}%
\bibitem [{\citenamefont {Wang}\ \emph {et~al.}(2010)\citenamefont {Wang},
  \citenamefont {Lv}, \citenamefont {Zhu},\ and\ \citenamefont {Ma}}]{glopt4}%
  \BibitemOpen
  \bibfield  {author} {\bibinfo {author} {\bibfnamefont {Y.}~\bibnamefont
  {Wang}}, \bibinfo {author} {\bibfnamefont {J.}~\bibnamefont {Lv}}, \bibinfo
  {author} {\bibfnamefont {L.}~\bibnamefont {Zhu}}, \ and\ \bibinfo {author}
  {\bibfnamefont {Y.}~\bibnamefont {Ma}},\ }\href {\doibase
  10.1103/PhysRevB.82.094116} {\bibfield  {journal} {\bibinfo  {journal} {Phys.
  Rev. B}\ }\textbf {\bibinfo {volume} {82}},\ \bibinfo {pages} {094116}
  (\bibinfo {year} {2010})}\BibitemShut {NoStop}%
\bibitem [{\citenamefont {Zhang}\ and\ \citenamefont {Buch}(1990)}]{lowestu2}%
  \BibitemOpen
  \bibfield  {author} {\bibinfo {author} {\bibfnamefont {Q.}~\bibnamefont
  {Zhang}}\ and\ \bibinfo {author} {\bibfnamefont {V.}~\bibnamefont {Buch}},\
  }\href {\doibase 10.1063/1.458536} {\bibfield  {journal} {\bibinfo  {journal}
  {J. Chem. Phys.}\ }\textbf {\bibinfo {volume} {92}},\ \bibinfo {pages} {5004}
  (\bibinfo {year} {1990})}\BibitemShut {NoStop}%
\bibitem [{\citenamefont {Ye}\ \emph {et~al.}(2009)\citenamefont {Ye},
  \citenamefont {Ming}, \citenamefont {Hu},\ and\ \citenamefont
  {Ning}}]{lowestu1}%
  \BibitemOpen
  \bibfield  {author} {\bibinfo {author} {\bibfnamefont {X.-X.}\ \bibnamefont
  {Ye}}, \bibinfo {author} {\bibfnamefont {C.}~\bibnamefont {Ming}}, \bibinfo
  {author} {\bibfnamefont {Y.-C.}\ \bibnamefont {Hu}}, \ and\ \bibinfo {author}
  {\bibfnamefont {X.-J.}\ \bibnamefont {Ning}},\ }\href {\doibase
  10.1063/1.3123042} {\bibfield  {journal} {\bibinfo  {journal} {J. Chem.
  Phys.}\ }\textbf {\bibinfo {volume} {130}},\ \bibinfo {pages} {164711}
  (\bibinfo {year} {2009})}\BibitemShut {NoStop}%
\bibitem [{\citenamefont {kroto}\ \emph {et~al.}(1985)\citenamefont {kroto},
  \citenamefont {Heath}, \citenamefont {Brien}, \citenamefont {Curl},\ and\
  \citenamefont {Smalley}}]{c60}%
  \BibitemOpen
  \bibfield  {author} {\bibinfo {author} {\bibfnamefont {H.~W.}\ \bibnamefont
  {kroto}}, \bibinfo {author} {\bibfnamefont {J.~R.}\ \bibnamefont {Heath}},
  \bibinfo {author} {\bibfnamefont {S.~C.}\ \bibnamefont {Brien}}, \bibinfo
  {author} {\bibfnamefont {R.~F.}\ \bibnamefont {Curl}}, \ and\ \bibinfo
  {author} {\bibfnamefont {R.~E.}\ \bibnamefont {Smalley}},\ }\href@noop {}
  {\bibfield  {journal} {\bibinfo  {journal} {Nature}\ }\textbf {\bibinfo
  {volume} {318}},\ \bibinfo {pages} {162} (\bibinfo {year}
  {1985})}\BibitemShut {NoStop}%
\bibitem [{\citenamefont {Heggie}\ \emph {et~al.}(2016)\citenamefont {Heggie},
  \citenamefont {L.}, \citenamefont {Latham},\ and\ \citenamefont
  {Trevethan}}]{c60sw1}%
  \BibitemOpen
  \bibfield  {author} {\bibinfo {author} {\bibfnamefont {M.~I.}\ \bibnamefont
  {Heggie}}, \bibinfo {author} {\bibfnamefont {H.~G.}\ \bibnamefont {L.}},
  \bibinfo {author} {\bibfnamefont {C.~D.}\ \bibnamefont {Latham}}, \ and\
  \bibinfo {author} {\bibfnamefont {T.}~\bibnamefont {Trevethan}},\ }\href@noop
  {} {\bibfield  {journal} {\bibinfo  {journal} {Phil. Trans. R. Soc. A}\
  }\textbf {\bibinfo {volume} {374:20150317}} (\bibinfo {year}
  {2016})}\BibitemShut {NoStop}%
\bibitem [{\citenamefont {Zhao}\ \emph {et~al.}(2003)\citenamefont {Zhao},
  \citenamefont {Lin},\ and\ \citenamefont {Yakobson}}]{c60sw2}%
  \BibitemOpen
  \bibfield  {author} {\bibinfo {author} {\bibfnamefont {Y.}~\bibnamefont
  {Zhao}}, \bibinfo {author} {\bibfnamefont {Y.}~\bibnamefont {Lin}}, \ and\
  \bibinfo {author} {\bibfnamefont {B.~I.}\ \bibnamefont {Yakobson}},\
  }\href@noop {} {\bibfield  {journal} {\bibinfo  {journal} {Phys. Rev. B}\
  }\textbf {\bibinfo {volume} {68}},\ \bibinfo {pages} {233403} (\bibinfo
  {year} {2003})}\BibitemShut {NoStop}%
\bibitem [{\citenamefont {Bettinger}\ \emph {et~al.}(2003)\citenamefont
  {Bettinger}, \citenamefont {Yakobson},\ and\ \citenamefont
  {Scuseria}}]{c60sw3}%
  \BibitemOpen
  \bibfield  {author} {\bibinfo {author} {\bibfnamefont {H.~F.}\ \bibnamefont
  {Bettinger}}, \bibinfo {author} {\bibfnamefont {B.~I.}\ \bibnamefont
  {Yakobson}}, \ and\ \bibinfo {author} {\bibfnamefont {G.~E.}\ \bibnamefont
  {Scuseria}},\ }\href@noop {} {\bibfield  {journal} {\bibinfo  {journal} {J.
  Am. Chem. Soc.}\ }\textbf {\bibinfo {volume} {125}},\ \bibinfo {pages} {5572}
  (\bibinfo {year} {2003})}\BibitemShut {NoStop}%
\bibitem [{\citenamefont {Li}\ and\ \citenamefont {Ning}(2004)}]{c60swnxj}%
  \BibitemOpen
  \bibfield  {author} {\bibinfo {author} {\bibfnamefont {P.}~\bibnamefont
  {Li}}\ and\ \bibinfo {author} {\bibfnamefont {X.-J.}\ \bibnamefont {Ning}},\
  }\href@noop {} {\bibfield  {journal} {\bibinfo  {journal} {J. Chem. Phys.}\
  }\textbf {\bibinfo {volume} {121}},\ \bibinfo {pages} {7701} (\bibinfo {year}
  {2004})}\BibitemShut {NoStop}%
\bibitem [{\citenamefont {Evans}\ and\ \citenamefont
  {Holian}(1985)}]{nosehoover}%
  \BibitemOpen
  \bibfield  {author} {\bibinfo {author} {\bibfnamefont {D.~J.}\ \bibnamefont
  {Evans}}\ and\ \bibinfo {author} {\bibfnamefont {B.~L.}\ \bibnamefont
  {Holian}},\ }\href@noop {} {\bibfield  {journal} {\bibinfo  {journal} {J.
  Chem. Phys.}\ }\textbf {\bibinfo {volume} {83}},\ \bibinfo {pages} {4069}
  (\bibinfo {year} {1985})}\BibitemShut {NoStop}%
\bibitem [{\citenamefont {Plimpton}(1995)}]{lammps}%
  \BibitemOpen
  \bibfield  {author} {\bibinfo {author} {\bibfnamefont {S.}~\bibnamefont
  {Plimpton}},\ }\href@noop {} {\bibfield  {journal} {\bibinfo  {journal}
  {Journal of Computational Physics}\ }\textbf {\bibinfo {volume} {117}},\
  \bibinfo {pages} {1} (\bibinfo {year} {1995})}\BibitemShut {NoStop}%
\bibitem [{\citenamefont {Brenner}\ \emph {et~al.}(2002)\citenamefont
  {Brenner}, \citenamefont {Shenderova}, \citenamefont {Harrison},
  \citenamefont {Stuart}, \citenamefont {Ni},\ and\ \citenamefont
  {Sinnott}}]{brenner}%
  \BibitemOpen
  \bibfield  {author} {\bibinfo {author} {\bibfnamefont {D.}~\bibnamefont
  {Brenner}}, \bibinfo {author} {\bibfnamefont {O.}~\bibnamefont {Shenderova}},
  \bibinfo {author} {\bibfnamefont {J.}~\bibnamefont {Harrison}}, \bibinfo
  {author} {\bibfnamefont {S.}~\bibnamefont {Stuart}}, \bibinfo {author}
  {\bibfnamefont {B.}~\bibnamefont {Ni}}, \ and\ \bibinfo {author}
  {\bibfnamefont {S.}~\bibnamefont {Sinnott}},\ }\href@noop {} {\bibfield
  {journal} {\bibinfo  {journal} {J. Phys.: Condensed Matter}\ }\textbf
  {\bibinfo {volume} {14}},\ \bibinfo {pages} {783} (\bibinfo {year}
  {2002})}\BibitemShut {NoStop}%
\bibitem [{\citenamefont {Cleri}\ and\ \citenamefont {Rosato}(1993)}]{tb}%
  \BibitemOpen
  \bibfield  {author} {\bibinfo {author} {\bibfnamefont {F.}~\bibnamefont
  {Cleri}}\ and\ \bibinfo {author} {\bibfnamefont {V.}~\bibnamefont {Rosato}},\
  }\href {\doibase 10.1103/PhysRevB.48.22} {\bibfield  {journal} {\bibinfo
  {journal} {Phys. Rev. B}\ }\textbf {\bibinfo {volume} {48}},\ \bibinfo
  {pages} {22} (\bibinfo {year} {1993})}\BibitemShut {NoStop}%
\bibitem [{\citenamefont {Verlet}(1967)}]{verlet}%
  \BibitemOpen
  \bibfield  {author} {\bibinfo {author} {\bibfnamefont {L.}~\bibnamefont
  {Verlet}},\ }\href {\doibase 10.1103/PhysRev.159.98} {\bibfield  {journal}
  {\bibinfo  {journal} {Phys. Rev.}\ }\textbf {\bibinfo {volume} {159}},\
  \bibinfo {pages} {98} (\bibinfo {year} {1967})}\BibitemShut {NoStop}%
\bibitem [{\citenamefont {Louwerse}\ and\ \citenamefont
  {Baerends}(2006)}]{vt1}%
  \BibitemOpen
  \bibfield  {author} {\bibinfo {author} {\bibfnamefont {M.~J.}\ \bibnamefont
  {Louwerse}}\ and\ \bibinfo {author} {\bibfnamefont {E.~J.}\ \bibnamefont
  {Baerends}},\ }\href@noop {} {\bibfield  {journal} {\bibinfo  {journal}
  {Chem. Phys. Lett.}\ }\textbf {\bibinfo {volume} {421}},\ \bibinfo {pages}
  {138 } (\bibinfo {year} {2006})}\BibitemShut {NoStop}%
\bibitem [{\citenamefont {Thompson}\ \emph {et~al.}(2009)\citenamefont
  {Thompson}, \citenamefont {Plimpton},\ and\ \citenamefont {Mattson}}]{vt2}%
  \BibitemOpen
  \bibfield  {author} {\bibinfo {author} {\bibfnamefont {A.~P.}\ \bibnamefont
  {Thompson}}, \bibinfo {author} {\bibfnamefont {S.~J.}\ \bibnamefont
  {Plimpton}}, \ and\ \bibinfo {author} {\bibfnamefont {W.}~\bibnamefont
  {Mattson}},\ }\href {\doibase 10.1063/1.3245303} {\bibfield  {journal}
  {\bibinfo  {journal} {J. Chem. Phys.}\ }\textbf {\bibinfo {volume} {131}},\
  \bibinfo {pages} {154107} (\bibinfo {year} {2009})}\BibitemShut {NoStop}%
\bibitem [{\citenamefont {Tsai}(1979)}]{vt3}%
  \BibitemOpen
  \bibfield  {author} {\bibinfo {author} {\bibfnamefont {D.~H.}\ \bibnamefont
  {Tsai}},\ }\href {\doibase 10.1063/1.437577} {\bibfield  {journal} {\bibinfo
  {journal} {J. Chem. Phys.}\ }\textbf {\bibinfo {volume} {70}},\ \bibinfo
  {pages} {1375} (\bibinfo {year} {1979})}\BibitemShut {NoStop}%
\bibitem [{\citenamefont {Agrawal}\ \emph {et~al.}(2002)\citenamefont
  {Agrawal}, \citenamefont {Rice},\ and\ \citenamefont
  {Thompson}}]{LJpotential}%
  \BibitemOpen
  \bibfield  {author} {\bibinfo {author} {\bibfnamefont {P.~M.}\ \bibnamefont
  {Agrawal}}, \bibinfo {author} {\bibfnamefont {B.~M.}\ \bibnamefont {Rice}}, \
  and\ \bibinfo {author} {\bibfnamefont {D.~L.}\ \bibnamefont {Thompson}},\
  }\href@noop {} {\bibfield  {journal} {\bibinfo  {journal} {Surf. Sci.}\
  }\textbf {\bibinfo {volume} {515}},\ \bibinfo {pages} {21 } (\bibinfo {year}
  {2002})}\BibitemShut {NoStop}%
\bibitem [{\citenamefont {Kraus}\ \emph {et~al.}(2016)\citenamefont {Kraus},
  \citenamefont {Davis}, \citenamefont {Seagle}, \citenamefont {Fratanduono},
  \citenamefont {Swift}, \citenamefont {Brown},\ and\ \citenamefont
  {Eggert}}]{copper1}%
  \BibitemOpen
  \bibfield  {author} {\bibinfo {author} {\bibfnamefont {R.~G.}\ \bibnamefont
  {Kraus}}, \bibinfo {author} {\bibfnamefont {J.-P.}\ \bibnamefont {Davis}},
  \bibinfo {author} {\bibfnamefont {C.~T.}\ \bibnamefont {Seagle}}, \bibinfo
  {author} {\bibfnamefont {D.~E.}\ \bibnamefont {Fratanduono}}, \bibinfo
  {author} {\bibfnamefont {D.~C.}\ \bibnamefont {Swift}}, \bibinfo {author}
  {\bibfnamefont {J.~L.}\ \bibnamefont {Brown}}, \ and\ \bibinfo {author}
  {\bibfnamefont {J.~H.}\ \bibnamefont {Eggert}},\ }\href {\doibase
  10.1103/PhysRevB.93.134105} {\bibfield  {journal} {\bibinfo  {journal} {Phys.
  Rev. B}\ }\textbf {\bibinfo {volume} {93}},\ \bibinfo {pages} {134105}
  (\bibinfo {year} {2016})}\BibitemShut {NoStop}%
\bibitem [{\citenamefont {Dewaele}\ \emph {et~al.}(2004)\citenamefont
  {Dewaele}, \citenamefont {Loubeyre},\ and\ \citenamefont
  {Mezouar}}]{copper2}%
  \BibitemOpen
  \bibfield  {author} {\bibinfo {author} {\bibfnamefont {A.}~\bibnamefont
  {Dewaele}}, \bibinfo {author} {\bibfnamefont {P.}~\bibnamefont {Loubeyre}}, \
  and\ \bibinfo {author} {\bibfnamefont {M.}~\bibnamefont {Mezouar}},\ }\href
  {\doibase 10.1103/PhysRevB.70.094112} {\bibfield  {journal} {\bibinfo
  {journal} {Phys. Rev. B}\ }\textbf {\bibinfo {volume} {70}},\ \bibinfo
  {pages} {094112} (\bibinfo {year} {2004})}\BibitemShut {NoStop}%
\bibitem [{\citenamefont {Tegeler}\ \emph {et~al.}(1999)\citenamefont
  {Tegeler}, \citenamefont {Span},\ and\ \citenamefont {Wagner}}]{areos}%
  \BibitemOpen
  \bibfield  {author} {\bibinfo {author} {\bibfnamefont {C.}~\bibnamefont
  {Tegeler}}, \bibinfo {author} {\bibfnamefont {R.}~\bibnamefont {Span}}, \
  and\ \bibinfo {author} {\bibfnamefont {W.}~\bibnamefont {Wagner}},\
  }\href@noop {} {\bibfield  {journal} {\bibinfo  {journal} {J. Phys. Chem.
  Ref. Data}\ }\textbf {\bibinfo {volume} {28}},\ \bibinfo {pages} {779}
  (\bibinfo {year} {1999})}\BibitemShut {NoStop}%
\bibitem [{\citenamefont {Young}\ \emph {et~al.}(2016)\citenamefont {Young},
  \citenamefont {Cynn}, \citenamefont {S\"{o}derlind},\ and\ \citenamefont
  {Landa}}]{areos2}%
  \BibitemOpen
  \bibfield  {author} {\bibinfo {author} {\bibfnamefont {D.~A.}\ \bibnamefont
  {Young}}, \bibinfo {author} {\bibfnamefont {H.}~\bibnamefont {Cynn}},
  \bibinfo {author} {\bibfnamefont {P.}~\bibnamefont {S\"{o}derlind}}, \ and\
  \bibinfo {author} {\bibfnamefont {A.}~\bibnamefont {Landa}},\ }\href
  {\doibase 10.1063/1.4963086} {\bibfield  {journal} {\bibinfo  {journal} {J.
  Phys. Chem. Ref. Data}\ }\textbf {\bibinfo {volume} {45}},\ \bibinfo {pages}
  {043101} (\bibinfo {year} {2016})}\BibitemShut {NoStop}%
\bibitem [{\citenamefont {Desgranges}\ and\ \citenamefont
  {Delhommelle}(2012)}]{arwl}%
  \BibitemOpen
  \bibfield  {author} {\bibinfo {author} {\bibfnamefont {C.}~\bibnamefont
  {Desgranges}}\ and\ \bibinfo {author} {\bibfnamefont {J.}~\bibnamefont
  {Delhommelle}},\ }\href {\doibase 10.1063/1.4712023} {\bibfield  {journal}
  {\bibinfo  {journal} {J. Chem. Phys.}\ }\textbf {\bibinfo {volume} {136}},\
  \bibinfo {pages} {184107} (\bibinfo {year} {2012})}\BibitemShut {NoStop}%
\bibitem [{\citenamefont {Lewis}\ \emph {et~al.}(1974)\citenamefont {Lewis},
  \citenamefont {Benson}, \citenamefont {Crawford},\ and\ \citenamefont
  {Daniels}}]{isochoric}%
  \BibitemOpen
  \bibfield  {author} {\bibinfo {author} {\bibfnamefont {W.}~\bibnamefont
  {Lewis}}, \bibinfo {author} {\bibfnamefont {D.}~\bibnamefont {Benson}},
  \bibinfo {author} {\bibfnamefont {R.}~\bibnamefont {Crawford}}, \ and\
  \bibinfo {author} {\bibfnamefont {W.}~\bibnamefont {Daniels}},\ }\href@noop
  {} {\bibfield  {journal} {\bibinfo  {journal} {J. Phys. Chem. Solids}\
  }\textbf {\bibinfo {volume} {35}},\ \bibinfo {pages} {383 } (\bibinfo {year}
  {1974})}\BibitemShut {NoStop}%
\bibitem [{\citenamefont {Gong}\ \emph {et~al.}(2019)\citenamefont {Gong},
  \citenamefont {Ning}, \citenamefont {Weng},\ and\ \citenamefont
  {Ning}}]{glcarxiv}%
  \BibitemOpen
  \bibfield  {author} {\bibinfo {author} {\bibfnamefont {L.-C.}\ \bibnamefont
  {Gong}}, \bibinfo {author} {\bibfnamefont {B.-Y.}\ \bibnamefont {Ning}},
  \bibinfo {author} {\bibfnamefont {T.-C.}\ \bibnamefont {Weng}}, \ and\
  \bibinfo {author} {\bibfnamefont {X.-J.}\ \bibnamefont {Ning}},\ }\href@noop
  {} {\bibfield  {journal} {\bibinfo  {journal} {arXiv}\ ,\ \bibinfo {pages}
  {1902.07388}} (\bibinfo {year} {2019})}\BibitemShut {NoStop}%
\bibitem [{\citenamefont {Liu}\ \emph {et~al.}(2019{\natexlab{a}})\citenamefont
  {Liu}, \citenamefont {Ning}, \citenamefont {Gong}, \citenamefont {Weng},\
  and\ \citenamefont {Ning}}]{lyparxiv}%
  \BibitemOpen
  \bibfield  {author} {\bibinfo {author} {\bibfnamefont {Y.-P.}\ \bibnamefont
  {Liu}}, \bibinfo {author} {\bibfnamefont {B.-Y.}\ \bibnamefont {Ning}},
  \bibinfo {author} {\bibfnamefont {L.-C.}\ \bibnamefont {Gong}}, \bibinfo
  {author} {\bibfnamefont {T.-C.}\ \bibnamefont {Weng}}, \ and\ \bibinfo
  {author} {\bibfnamefont {X.-J.}\ \bibnamefont {Ning}},\ }\href@noop {}
  {\bibfield  {journal} {\bibinfo  {journal} {arXiv}\ ,\ \bibinfo {pages}
  {1901.09205}} (\bibinfo {year} {2019}{\natexlab{a}})}\BibitemShut {NoStop}%
\bibitem [{\citenamefont {Liu}\ \emph {et~al.}(2019{\natexlab{b}})\citenamefont
  {Liu}, \citenamefont {Ning}, \citenamefont {Gong}, \citenamefont {Weng},\
  and\ \citenamefont {Ning}}]{lyparxiv2}%
  \BibitemOpen
  \bibfield  {author} {\bibinfo {author} {\bibfnamefont {Y.-P.}\ \bibnamefont
  {Liu}}, \bibinfo {author} {\bibfnamefont {B.-Y.}\ \bibnamefont {Ning}},
  \bibinfo {author} {\bibfnamefont {L.-C.}\ \bibnamefont {Gong}}, \bibinfo
  {author} {\bibfnamefont {T.-C.}\ \bibnamefont {Weng}}, \ and\ \bibinfo
  {author} {\bibfnamefont {X.-J.}\ \bibnamefont {Ning}},\ }\href@noop {}
  {\bibfield  {journal} {\bibinfo  {journal} {arXiv}\ ,\ \bibinfo {pages}
  {1902.06248}} (\bibinfo {year} {2019}{\natexlab{b}})}\BibitemShut {NoStop}%
\end{thebibliography}
%

\end{document}